\documentclass{article}

\usepackage{amsmath}
\usepackage{amssymb}
\usepackage{amsthm}
\usepackage{amsmath}
\usepackage{mathtools}
\usepackage{bbm}
\usepackage[a4paper,top=3cm,bottom=3cm,left=3cm,right=3cm,  bindingoffset=5mm, marginparwidth=1.75cm]{geometry}
\usepackage[numbers]{natbib}
\usepackage{graphicx}
\usepackage{dsfont}
\usepackage{textcomp}
\usepackage{url}
\usepackage{sidecap}
\usepackage{array}
\usepackage{chngcntr}
\usepackage{MnSymbol}
\usepackage{wasysym}
\usepackage[hidelinks]{hyperref}
 
\usepackage{epstopdf}
\usepackage{framed}
\usepackage{xcolor} 
\usepackage{transparent}
\usepackage{blindtext}
\usepackage{float}
\usepackage{caption}
\usepackage{subcaption}
\usepackage{authblk}
\usepackage{natbib}
\usepackage{rotating}
\usepackage{nomencl}
\usepackage{paralist}
\usepackage{fancyhdr}
\usepackage{pgf}
\providecommand{\keywords}[1]{\textbf{\textbf{Keywords}} #1}

\newcommand{\Prob}{\mathbb{P}} 
\newcommand{\E}{\mathbb{E}} 
\newcommand{\Z}{\mathbb{Z}} 
\newcommand{\R}{\mathbb{R}} 
\newcommand{\N}{\mathbb{N}} 
\newcommand{\X}{\mathbb{X}} 
\newcommand{\St}{\mathbb{S}} 
\newcommand{\ke}{k^{\epsilon}} 
\newcommand{\ka}{\tilde{k}^{\epsilon} } 
\newcommand{\Bl}{\mathcal{B}} 
\newcommand{\1}{\mathbbm{1}} 
\newcommand{\C}{C} 
\newcommand{\trans}{P} 
\newcommand{\transback}{P^-} 

\usepackage{natbib}
\usepackage{graphicx}

\title{Statistical Analysis of Tipping Pathways in Agent-Based Models}

 \author[a,b,c]{Luzie Helfmann}
 \author[c]{Jobst Heitzig}
 \author[a]{P{\'e}ter Koltai}
 \author[c,d]{J{\"u}rgen Kurths}
 \author[a,b]{Christof Sch{\"u}tte}
 
 \affil[a]{Institute of Mathematics, Freie Universit\" at Berlin, Berlin,  Germany}
 \affil[b]{Zuse-Institute Berlin, Berlin, Germany}
 \affil[c]{Department of Complexity Science, Potsdam Institute for Climate Impact Research, Potsdam, Germany}
 \affil[d]{Department of Physics, Humboldt University, Berlin, Germany}

\date{March 2021}

\begin{document}
 
\maketitle
\begin{abstract}
Agent-based models are a natural choice for modeling complex social systems.
In such models simple stochastic interaction rules for a large population of individuals can lead to emergent dynamics on the macroscopic scale, for instance a sudden shift of majority opinion or behavior.
Here, we are concerned with studying noise-induced tipping between relevant subsets of the agent state space representing characteristic configurations.
Due to the large number of interacting individuals, agent-based models are high-dimensional, though usually a lower-dimensional structure of the emerging collective behaviour exists.
We therefore apply Diffusion Maps, a non-linear dimension reduction technique, to reveal the intrinsic low-dimensional structure. 
We characterize the tipping behaviour by means of Transition Path Theory, which helps gaining a statistical understanding of the tipping paths such as their distribution, flux and rate. 
By systematically studying two agent-based models that exhibit a multitude of tipping pathways and cascading effects, we illustrate the practicability of our approach. 
\end{abstract}


\section{Introduction}
Understanding tipping pathways and tipping cascades in  social systems is very important for our interconnected  world.
Tipping is defined as a qualitative change from one rather stable state to another one upon a small quantitative change, e.g., of a  parameter, or due to noise. One can distinguish between the following general types of tipping~\cite{ashwin2012tipping}: 
In \emph{bifurcation-induced} tipping,  the state of a parameter-dependent system changes qualitatively due to an external control parameter crossing a threshold value. The parameter is assumed to vary infinitesimally slowly such that one studies transitions of stationary dynamics. Often this threshold value is called the \emph{tipping point} or the \emph{point of no return}.
In \emph{noise-induced} tipping,  noisy fluctuations result in the system escaping from the neighborhood of a metastable state.
Last,  \emph{rate-induced} tipping happens  when the control parameter changes faster than a certain critical rate of change such that the system fails to track the continuously changing attractor.

Social systems are complex systems and as such characterized  by   rich, nonlinear and usually local (i.e., only between neighbours) interactions among a large number of individual constituents~\cite{cilliers1999complexity,cilliers2001boundaries}. The individual entities are ignorant of the behaviour of the system as a whole  and only respond to local information. Social systems are usually open, i.e.,  continuously in interaction with their environment, and  therefore often not in a simple equilibrium. Moreover, the history of the system affects the present. Hierarchies and multi-scale structures are present in complex social systems~\cite{simon1991architecture}.
When modeling social systems, agent-based models (ABMs) are a natural choice.
One defines the characteristics of a large sample of discrete entities, the so-called agents (e.g., people, animals, cars, companies, \dots), and a set of possible actions and local interactions rules for the agents. Often these are stochastic, thus reflecting the unpredictability and individuality of the agents. 
From the interplay of local interactions, global patterns can emerge~\cite{macy2002factors}.

The complexity of social systems is inherited by their tipping dynamics. 
In~\cite{winkelmann2020social}, it is argued that tipping in social systems, such as an epidemic, a social contagion process of ideas or norms~\cite{gladwell2006tipping, nyborg2016social}, a crash in a financial system, or 
the diffusion of a new technology~\cite{sharpe2020upward},  faces several difficulties and can be much more complex than tipping in climate or ecological systems. ``Tipping in realistic social systems can usually not be linked to a single control parameter, instead multiple interrelated factors act as a forcing of the transitions (e.g. policies, communication, taxation, \dots). Moreover, there is a larger number of mechanisms that cause tipping and various pathways of change towards a greater number of potentially stable post-tipping states."~\cite{winkelmann2020social}
Recently, there has been an increasing interest in studying \emph{tipping cascades}, i.e., cascading effects in interacting systems where the tipping of one sub-system influences the tipping likelihood of another one. Interactions between tipping elements have been studied in the climate system~\cite{kriegler2009imprecise,cai2016risk}, in ecological systems~\cite{sahasrabudhe2011rescuing}, and also in social systems~\cite{sharpe2020upward}.

In this paper we will analyse noise-induced tipping in certain high-dimensional agent-based models. We will assume that the system is stationary, though we comment on a generalization to non-stationary systems, where for example noise-induced tipping can happen in combination with bifurcation-induced tipping.

The idea of our approach is to first reduce the dimension of the ABM by means of a nonlinear dimension reduction technique, Diffusion Maps, thereby relying on the existence of some lower-dimensional structure on which the ABM dynamics essentially takes place. This assumption can be made for most ABMs, since the emergence of macro-scale patterns and collective behaviour is a key property of ABMs. Central to this is that we do not need to know which macroscopic features the system eventually evolves along, but we can learn the associated coordinates from sufficiently rich dynamical data~\cite{kemeth2020learning}.
Diffusion Maps have already been applied for finding low-dimensional coordinates, so-called  \emph{collective variables}, \emph{reaction coordinates}, or \emph{order parameters}, of ABMs~\cite{liu2014coarse,marschler2014coarse}.

Since tipping in stochastic ABMs is characterized by the existence of many meta\-stable states and a multitude of transition pathways between them, we will apply Transition Path Theory (TPT), and thereby gain a complete statistical understanding of the ensemble of transition paths between two chosen  subsets $A$ and $B$ of the state space~\cite{weinan2006towards, Vanden-Eijnden2006,metzner2009transition}. 
For studying noise-induced tipping between two metastabilities, one chooses $A$ and $B$ as metastable sets, i.e., two sets in which the system is trapped for a comparatively long time, but can eventually also escape from them. But TPT does not restrict $A$ and $B$ to be metastable, one can also study transitions between other relevant subsets of the state space, for instance given by viability constraints.  
TPT builds on the information  that is contained  in the \emph{forward} and \emph{backward committor functions}, i.e., in the hitting probability of $B$ forward in time and of $A$ backward in time. The advantage of studying tipping by TPT is that it allows to unravel the full range of transition pathways between    sets $A$ and $B$ by computing the flux of transition paths, as well as other statistical properties of the transition paths, e.g., their distribution, rate and mean duration.

Recently, the forward committor has been singled out as the central object for quantifying the risk of future tipping~\cite{lucente2019machine,finkel2021learning}. Several papers study how one can solve for high-dimensional committors using neural networks~\cite{khoo2019solving,li2019computing,lucente2019machine,li2020solving}.
Very much related to tipping is the concept of resilience, which in its simplest form is the system's ability of returning to the original state or region after a perturbation. Using similar objects as in TPT, namely escape probabilities and committors,  this form of resilience of a system when in some or other attractor can be studied by analysing their stochastic basin of attraction~\cite{serdukova2016stochastic, serdukova2017metastability, lindner2019stochastic}. 

TPT was originally developed  for  studying rare transitions in statistical mechanics, e.g., protein folding~\cite{noe2009constructing}, and chemical reactions, but was later also applied for analysing transition events in the climate system~\cite{finkel2020path} and marine debris dynamics~\cite{miron2020transition}. 
Bifurcation diagrams of high-dimensional ABMs have already been studied~\cite{tsoumanis2010equation, siettos2012equation}, as well as the various bifurcation-induced transition pathways in a coupled social-ecological model~\cite{mathias2020exploring}, but to our knowledge noise-induced tipping in agent-based models has not been considered yet.  

We will illustrate our approach on two paradigmatic models that exhibit tipping. The first model is based on Granovetter's threshold model~\cite{granovetter1978threshold} and describes the social activation of people for  some collective  action, such as rioting. Therein, when at least a certain fraction of an agents' neighbourhood is active in the collective action, the agent has a high chance of also becoming active.
The second model considers agents that influence each other regarding their opinions and actual behavioural choices with respect to certain behavioural options, such as a more or less climate-friendly lifestyle or following certain epidemic countermeasures more or less stringently. That model assumes that the more agents in the population hold the opinion that ``one should do A'' (e.g., wear a face mask), the more likely an agent can be convinced by her social peers to switch from choosing behaviour non-A to behaviour A for themselves. At the same time, the more agents in the population actually exhibit behaviour A, the more likely an agent can be convinced by her social peers to switch from the opinion ``one should do A'' to ``one should not do A'', since it may seem that the issue addressed by behaviour A is already sufficiently dealt with. This negative feedback loop then induces oscillatory dynamics. 
We will study both models on highly modular interaction networks, where the different blocks of agents (i.e., densely connected groups) influence each other. When one block changes its state, i.e., ``tips", connected blocks are more likely to also change.  Thus tipping happens as a tipping cascade among connected blocks.

Our overall approach has the perspective of giving a quantitative analysis of noise-induced tipping  without making any prior structural assumptions about the system. 
Instead of studying tipping points in bistable systems or the stability of the attractors, 
Transition Path Theory offers a new perspective onto tipping by quantitatively characterising the dominant pathways along which tipping happens. This allows for a more detailed understanding of the tipping process especially for complicated systems, as well as for finding new ways of bypassing and preventing tipping. 

In the following, we will first in Section~\ref{sec:models} introduce two agent-based models that exhibit noise-induced
tipping. In Section~\ref{sec:coll_var} we will show how we can find a reduced representation in terms of collective variables by using the Diffusion Maps algorithm. This allows us
to finally in Section~\ref{sec:tipping} analyse the tipping pathways using Transition Path Theory.

\section{Two agent-based models exhibiting tipping}\label{sec:models}
We start by introducing two agent-based models (ABMs) that exhibit tipping.
In these models, agents are making behavioural decisions and change their opinions in reaction to the social influence of their network neighbours, potentially mediated by an additional macroscopic interaction. 
Apart from their fixed position in the network, agents are identical.
We will assume interaction networks consisting of several groups of nodes which are densely linked among themselves but with only few connections to the other groups.
The densely linked agents in each block are nearly identical because they are connected to very similar sets of other agents, and thus behave rather similarly due to the local interaction rules. Thus both ABMs have many metastable states,  where agents behave collectively in each of the blocks. 

Many ABMs can be written as Markov chains or Markov jump processes, see~\cite{izquierdo2009techniques} for some examples.  The Markovianity assumption means that the next state of the system only depends on the current state and not the history.\footnote{This does not necessarily mean that agents have no memory or cannot be influenced by their past. By enlarging the state space formally to include a memory of past states, Markovian dynamics can be retained.} 
The two models that we consider can be viewed as Markov chains $(\mathbf{X}_t)_{t\in \Z}$.
They have a finite, but large state space and are irreducible, thus ergodic, as well as aperiodic.   
Due to these properties, both ABMs exhibit tipping, i.e., transitions between several metastable states. Later, we are interested in studying noise-induced tipping and tipping cascades between two diametrically opposed metastable regions of the dynamics. 
For the tipping analysis, we consider the models in stationarity.

For a comprehensive introduction to Markov chains we refer the reader to the book of Norris~\cite{norris1998markov}.
Note that we follow the convention to use uppercase letters $X$ for random variables and lowercase letters $x$ for their possible realizations.

\subsection{A threshold model of social contagion or activation}
We will introduce and discuss a very simple ABM of social contagion to describe phenomena such as the spreading of cultural fads, hypes or consumption behaviours, or the activation for some collective action such as rioting.

Let us consider a population of agents where each agent can be in one of two discrete states:  being \emph{inactive} or \emph{active} in the collective action.
The interaction topology between agents is given by a fixed network. 
A threshold-like influence is exerted by the social neighbours  when an agent makes a binary decision: if more than a certain fraction of neighbours are in the opposite state to that of the agent, the agent will switch its state with a high probability. Thus each agent aligns its state with the state of the majority of its social  neighbours. In addition, there is  a small probability for the agent to switch its state without social influence, which can either be interpreted as a form of exploration or as representing otherwise unmodelled additional causes for switching one's state.

This  ABM is ultimately based on Granovetter's threshold model, but Granovetter  considered a fully mixed population~\cite{granovetter1978threshold}. More recently, several network-based versions of his original idea have been proposed~\cite{watts2002simple, wiedermann2020network}, also containing different classes of agents such as stubborn agents that have a fixed state. Often, threshold distributions for the population are studied as well as deterministic interactions  resulting in only one  decision-making cascade through the population~\cite{granovetter1978threshold,watts2002simple,wiedermann2020network}. We instead consider the threshold to be constant for all agents and assign probabilities to the activity changes, thus  our system can escape from the metastable regions.

Let us define our threshold model in more detail: 

\paragraph{Interaction rules.}
We consider a system of $N$ interacting agents with social connections among them given by the edges of a static network $\mathcal{G}$ of $N$ vertices.   
The state of each agent $i$ at the discrete time point $t$ is denoted by $X^i_t\in \{0,1\}$ corresponding to being \emph{inactive} or \emph{active} in the collective action, respectively.

At each time $t=0,1,\dots$, each agent $i$ in state $X^i_t=0$ (resp.\ $1$) will change their state to $X^i_{t+1}=1$ (resp. $0$) 
\begin{itemize}
\item with probability  $p$, if more than or exactly a fraction $\theta$ of neighbouring agents at time $t$ are in the opposite state $1$  (resp. $0$), 
\item or with the exploration probability $e$, if less than a fraction $\theta$ of neighbours is in the opposite state,  
\end{itemize}
where we assume $1 > p\gg e >0$ such that social influence is stronger than exploration.
        
We can also view the system as a Markov chain $(\mathbf{X}_t)_{t\in \mathbb{Z}}$ on the state space \mbox{$\X = \{0,1\}^N$}, where we denote the \emph{population state} at time $t$ by $\mathbf{X}_t = (X^i_t)_{i = 1}^N$. Since agents in every time step change their state synchronously and independently of each other, the transition matrix on $\X$ decomposes into the product of the ``transition probabilities" for each individual agent
\begin{equation}
\trans(\mathbf{x},\mathbf{y}) := \Prob(\mathbf{X}_{t+1} = \mathbf{y} \mid \mathbf{X}_t = \mathbf{x}) = \prod_{i=1}^N \Prob(X^i_{t+1} = y^i \mid \mathbf{X}_t = \mathbf{x}).
\label{eq:T_decomp}
\end{equation}
The exploration probability ensures that agents are never stuck in a state. In every time step an agent has a positive probability to remain in the same state as well as to change the state,  i.e., $\Prob(X^i_{t+1} = 0 \mid\mathbf{X}_t = \mathbf{x})>0$ and $\Prob(X^i_{t+1} = 1 \mid\mathbf{X}_t = \mathbf{x})>0$ respectively.
Thus by \eqref{eq:T_decomp} there is a positive probability to go from any population state to any other within one time step, implying that the Markov chain is irreducible and also aperiodic. 
 
\paragraph{Interaction network.}
We assume that the interaction network $\mathcal{G}$ has two scales: it consists of \emph{blocks}, sometimes also referred to as \emph{communities} or \emph{clusters}, in which the nodes are densely connected, whereas nodes  of different blocks are sparsely connected.
One approach to randomly generate such a network is by the stochastic block model. Each node $i$ is assigned to a \emph{block} $\Bl_k$,~$k=1,\dots, K$. The blocks are disjoint subsets of the set of agents, thus when node $i$ belongs to $\Bl_k$, we write  $i\in \Bl_k$.  After defining a symmetric matrix $W = (W_{kl})$ of size $K\times K$ that contains the edge wiring probabilities between a node of block $\Bl_k$ and one of block $\Bl_l$, 
we go through all pairs of nodes independently and with probability $W_{kl}$ place an edge between them when they belong to blocks $\Bl_k$ and $\Bl_l$.
In the case of only one block, this is equivalent to the Erdős--Rényi random graph model. 
 
\paragraph{Resulting dynamics.}
\begin{SCfigure}
    \centering
    \includegraphics[width = 0.65\textwidth]{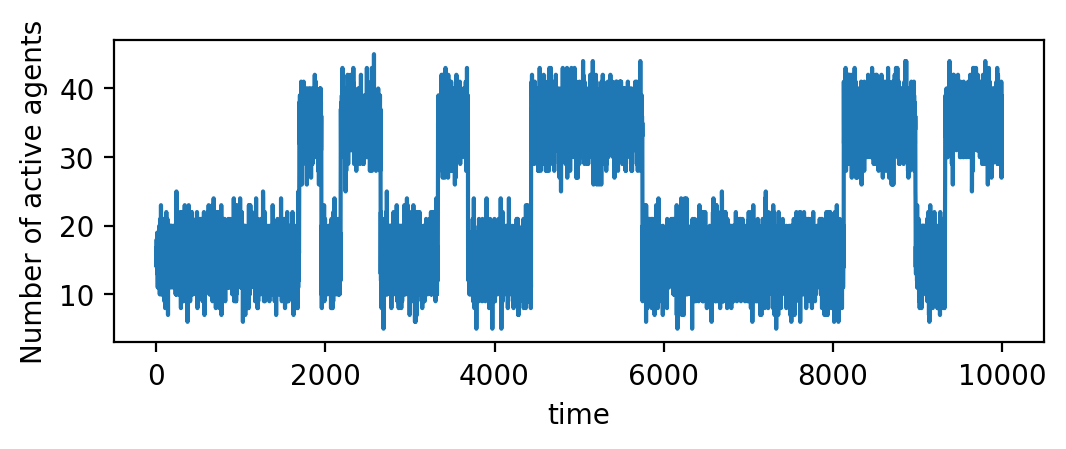}
    \caption{Realization of the threshold model with $50$ agents that are interacting on a complete network, i.e.,   each agent is influenced by the whole population. The dynamics switches between two metastable macrostates. The model parameters are $e=0.3$, $p=0.7$, $\theta=0.5$.
}
    \label{fig:grano_1cluster}
\end{SCfigure}
If we consider a population where every agent is interacting with every other agent, i.e., they are interacting on a complete network\footnote{We call a network complete if every node of the network is connected to every other node of the network.}, the system switches between two metastable regions: (i) where a majority of agents is inactive and (ii) where most agents are active, see Figure~\ref{fig:grano_1cluster} for a realization with $50$ agents.
If we now study a population consisting of several (complete or mostly complete) blocks, with some connections between the blocks, then this structure is multiplied. Each block can switch between two such metastable regions, but depending on the number of connections between the blocks, all blocks are either synchronized, only weakly influencing each other, or behaving mostly independently. 
We set parameters to the case where the blocks are weakly influencing each other in order to avoid a trivial behaviour and to focus on the most interesting dynamical regime.

\begin{figure}
     \centering
     \begin{subfigure}[b]{0.63\textwidth}
         \centering
         \includegraphics[width=1\textwidth]{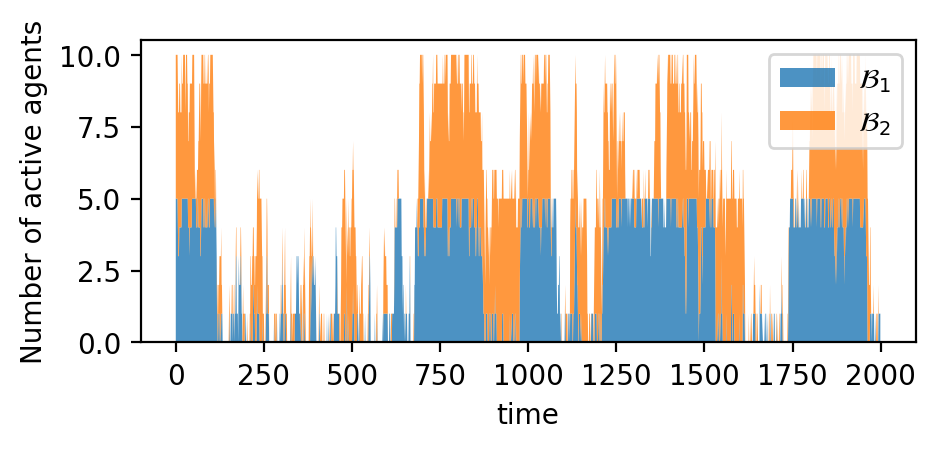}
         \caption{}
     \end{subfigure}
     \begin{subfigure}[b]{0.36\textwidth}
         \centering
         \includegraphics[width=\textwidth]{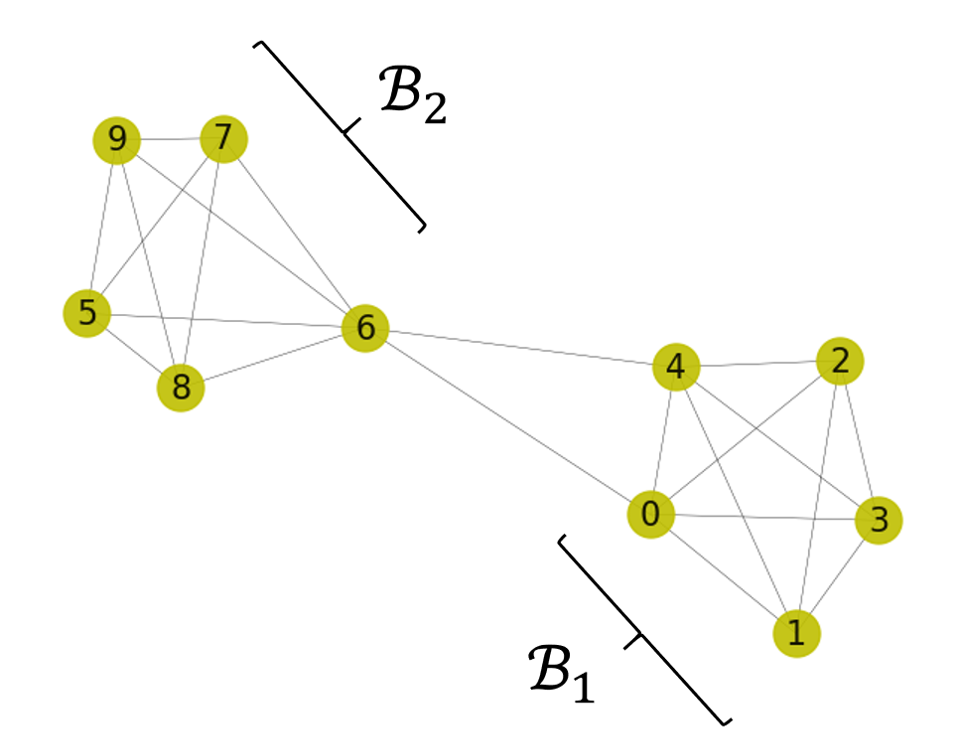}
         \caption{}
     \end{subfigure}     
     
     \begin{subfigure}[b]{1\textwidth}
         \centering
         \includegraphics[width=\textwidth]{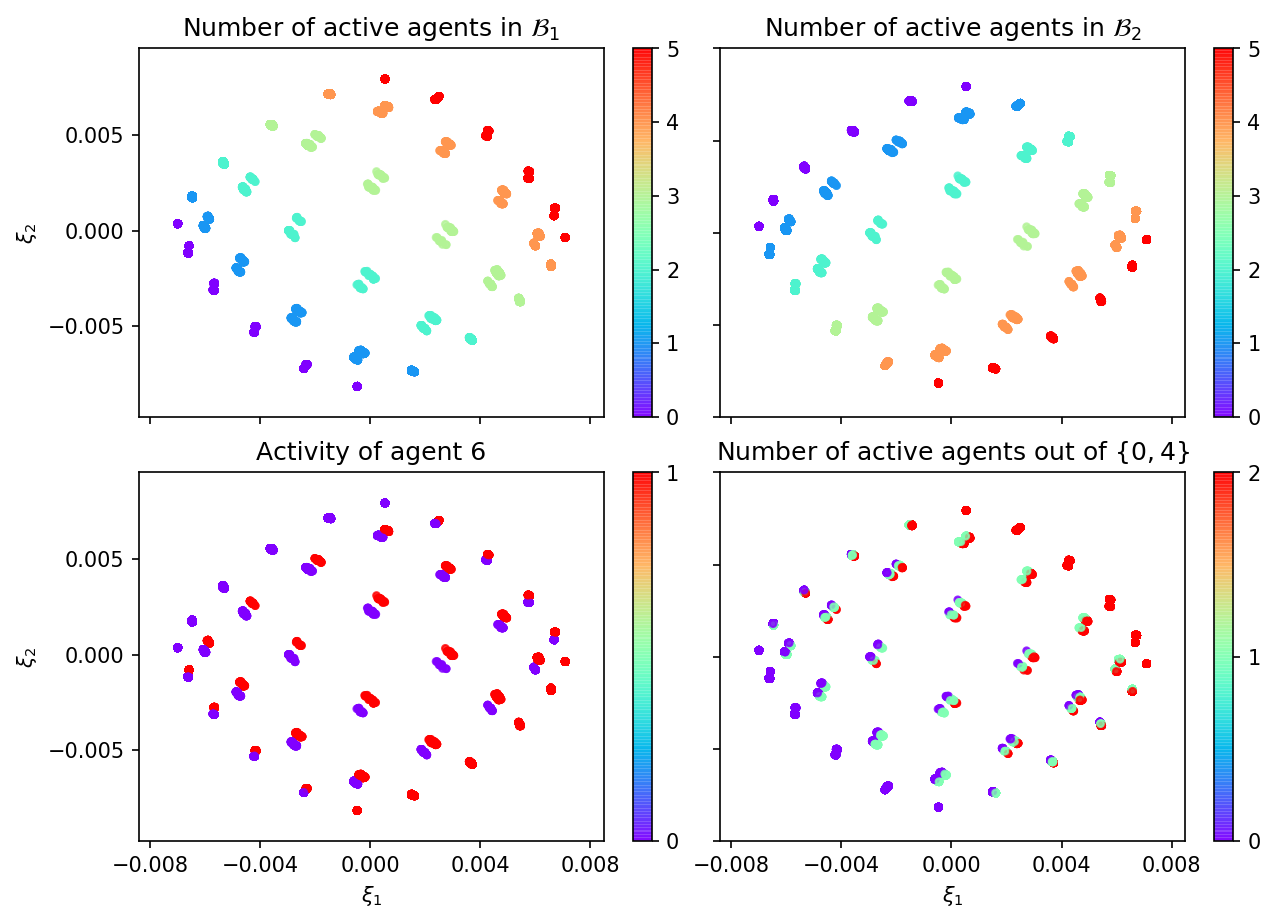}
         \caption{}
     \end{subfigure}

        \caption{Threshold model with two blocks of $5$ agents each as in \emph{Example 1}: (a) The realization is shown using a stackplot, i.e., the number of active agents in $\mathcal{B}_2$ is plotted vertically on top of the number of active agents in $\mathcal{B}_1$, (b) modular agent network, (c) projection of population states into the dominant two Diffusion Maps coordinates, the scale parameter turned out to be $\epsilon=0.25$, the data points are colored according to the number of active agents in each block and the activity of agents $0,4$ and $6$.
        }
        \label{fig:granovetter_2cluster}
\end{figure}

\textit{Example 1:}  The first example we will consider throughout the paper is a small population of just $10$ agents that are evenly split into two blocks. 
We set the change probability as $p=0.3$, the exploration probability as $e=0.03$ and the threshold $\theta=0.5$.
With a size of $|\X|=2^{10}$, the state space is already nontrivial, but still small enough to be able to do direct computations of the state space. 
In  Figure~\ref{fig:granovetter_2cluster}~(a) we show a realization on the small network that is shown in~\ref{fig:granovetter_2cluster}~(b). 
The realization indicates that the system remains in four metastable regions most of the time:  where  (i-ii) a majority in block 1 (resp. 2) but not in the other block is active, (iii) a majority in both blocks is inactive, and (iv) a majority in both blocks is active. 
It seems that those states (i-ii) where the two blocks show a different majority activity are less metastable than those states (iii-iv) where agents in both blocks are conform. 
Moreover, the realization suggests  that the tipping of one block induces the other block to also tip.
By ``tipping" we understand a transition from one metastable region to another, i.e., one block of agents drastically changes its state from majority of agents active to majority inactive (or vice versa). Sometimes we might refer to the tipping of the whole population, i.e., when all blocks drastically change between majority of agents being active and the majority being inactive. This happens via the individual blocks' successive tipping, i.e., via a \emph{tipping cascade}.

\begin{figure}
     \centering

     \begin{subfigure}[b]{1\textwidth}
         \centering
         \includegraphics[width=\textwidth]{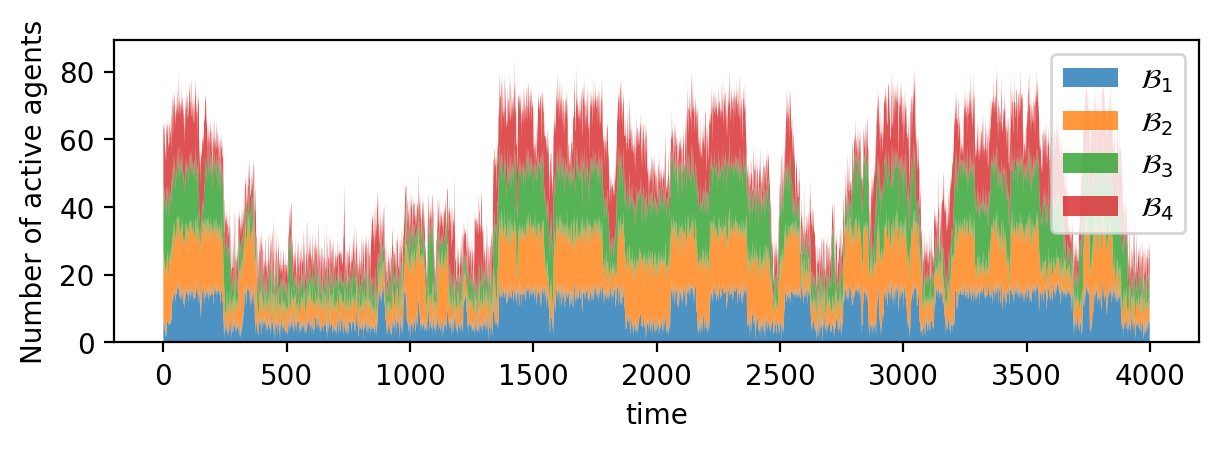}
         \caption{}
     \end{subfigure}

     \begin{subfigure}[b]{1\textwidth}
         \centering
         \includegraphics[width=\textwidth]{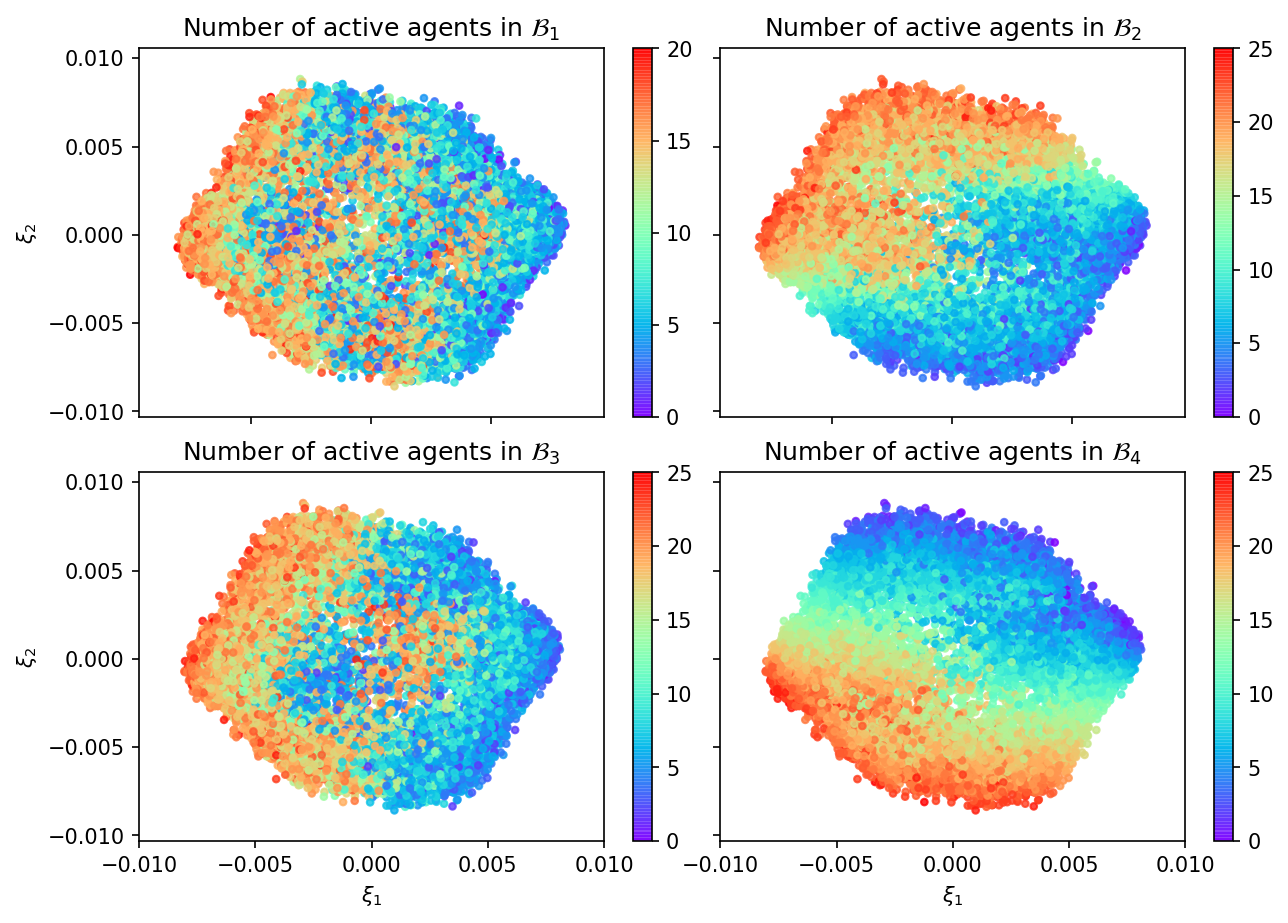}
         \caption{}
     \end{subfigure}
     
        \caption{Threshold model dynamics with four incomplete blocks as in \emph{Example 2:} (a) Realization shown as a stackplot, (b) Diffusion Maps projection into the first two coordinates and colored according to the number of active agents in each block, the Diffusion Maps scale parameter came out as $\epsilon=0.15$.}
        \label{fig:granovetter_4cluster}
\end{figure}

\textit{Example 2:} As a second example we consider a large population structured into four blocks of different sizes. Block 1 contains $20$ agents, all other blocks consist of $25$ agents, see Figure~\ref{fig:granovetter_4cluster_tpt}~(d) for the network. The four blocks are circularly connected, and the network is generated by the stochastic block model where each agent  has a wiring probability of $0.9$ to agents in the same block and of $0.04$ to agents from circularly neighboured blocks. We set $e=0.23$, $p=0.66$, $\theta=0.5$.
In this example there are potentially 16 metastable regions, since agents in each block can be mostly active or not and there are four blocks. 
One can again assume that the tipping of one block induces neighbouring blocks to also tip, see  Figure~\ref{fig:granovetter_4cluster}~(a) for a realization.

\subsection{An oscillating, bivariate complex contagion model}
In this second ABM we are modeling the changes of binary opinions and separately of  binary actual behavioural choices with respect to a certain behavioural option, such as a climate-friendly lifestyle or a certain preventive measure against an epidemic. For illustrative purposes, we use the context of climate-friendly lifestyles and the metaphor of ``green" behaviour. We hence say that each agent has a \emph{non-green} or \emph{green} opinion, and also displays a \emph{non-green} or \emph{green} actual behaviour.

The model again considers a complex contagion process~\cite{centola2007complex}, where the social reinforcement from multiple agents at the same time is needed for an agent to change its state. But this time an agent's state has two components: opinion and actual behavioural, and the model also does not have a sharp threshold-like rule. 
Instead, the state change in opinion resp.\ actual behaviour of an agent is triggered upon interacting with two neighbours that both hold the opposite opinion resp.\ both display the opposite behaviour. Additionally, the actual probabilities with which these switches then occur also depend on the macroscopic state of the agent's block. 
In the model a switch in an agent's actual behaviour is made more likely by the respective opinion in the agent's block (e.g., the more agents have a \emph{green} opinion, the more agents switch to a \emph{green} behaviour), whereas a change of opinion is amplified if the block displays the opposite behaviour (e.g., the more agents display a \emph{green} behaviour, the more agents will switch to a \emph{non-green} opinion), the resulting dynamics leads to oscillations, i.e., is cyclic. 

This model shows that opinions and actual behaviours do not always have to be aligned. There might be a time lag between holding a certain opinion and behaving accordingly.
Additionally, the incentive to hold a certain opinion drops when many agents in the block are behaving in that way. It seems that there is no longer the need to hold the respective opinion since enough action is taken by other agents.

In more detail the model is formulated as follows:
\paragraph{Setting.}
We consider a system of $N$ agents, each agent $i$ with a binary opinion \mbox{$O_t^i \in \{0,1\}$} and a binary behaviour $B_t^i \in \{0,1\}$ at time~$t$. For illustration we consider $0$ as \emph{non-green} and $1$ as \emph{green}.
In each time step, each agent is interacting with two randomly drawn neighbours in a static social network $\mathcal{G}$. We again assume an interaction network with many communities, e.g, generated by the stochastic block model, and that each agent has at least two neighbours. 
Further, each agent $i$ is influenced by the set of agents within the same block. For an agent  $i \in \Bl_l$, we define the  following \emph{block} fractions:
$$\bar{O}_t^i := \frac{|\{ j \in \Bl_l:  O_t^j=1\}|}{|\Bl_l|},$$
i.e.,  the fraction of agents with a \emph{green} opinion in the same block as $i$, and
$$\bar{B}_t^i := \frac{|\{ j \in \Bl_l:  B_t^j=1\}|}{|\Bl_l|},$$
the fraction of agents with a \emph{green} behaviour in the same block. Note that these quantities, viewed as functions of the agents' index $i$, are constant on each block.

Below, the parameters $b,c\in [0,1]$  determine how strongly a \emph{green} resp.\ \emph{non-green} change in behaviour is influenced by the opinions in the block.
Likewise, the parameters $f,g \in [0,1]$  determine how strongly a \emph{green} resp.\ \emph{non-green} change in opinion is influenced by the actual behaviour in the block. 
The general rate parameter $\tau \in (0,1)$ is for scaling the  amount of change per time step. 

\paragraph{Interaction rules.}
At each discrete time point $t$, each agent $i$ independently chooses two distinct neighbours $j,k$ uniformly at random.

A behaviour change occurs:
\begin{itemize}
    \item \begin{tabbing} if $B_t^j = B_t^k = 1$, $B_t^i=0$: \= agent $i$ changes its behaviour to $B_{t+1}^i=1$ \\
        \> with probability $\tau (b \, \bar{O}_t^i + (1-b))$, \end{tabbing}
    \item \begin{tabbing} if $B_t^j = B_t^k = 0$, $B_t^i=1$: \= agent $i$ changes its behaviour  to $B_{t+1}^i= 0$ \\
        \> with probability $\tau( c\,  (1 - \bar{O}_t^i) + (1-c)$), \end{tabbing}
    \item or else, with a small exploration probability $e$, agent $i$ changes its behaviour.
\end{itemize}
Thus an agent has a higher chance of changing its behaviour to  \emph{green}  when interacting with two neighbours of   \emph{green} behaviour  and the more likely the more agents in his block have a \emph{green} opinion.\footnote{Note that if we disregard the rate $\tau$, the first behaviour-change probability is a convex combination with factor $b$ between the probabilities $\bar{O}_t^i$ (``fraction in block with green opinion'') and~1 (``change with certainty to the behaviour of the two chosen neighbors'').}
An agent is more likely to change its behaviour to \emph{non-green}, when interacting with two  neighbours of \emph{non-green} behaviour  and the more agents  in his block show a \emph{non-green} behaviour. 

Conversely an opinion change happens:
\begin{itemize}
    \item \begin{tabbing} if $O_t^j = O_t^k = 1$, $O_t^i=0$: \= with probability $\tau  (f\, (1-\bar{B}_t^i) + (1-f))$,\\
        \> agent $i$ changes its opinion to $O_{t+1}^i=1$,\end{tabbing}
    \item \begin{tabbing} if $O_t^j = O_t^k = 0$, $O_t^i=1$: \= with probability $\tau (g\,   \bar{B}_t^i + (1-g) )$,\\
        \> agent $i$ changes its state $O_{t+1}^i=0$,\end{tabbing}
    \item or else: with a small probability $e$, agent $i$ changes its opinion.
\end{itemize}
This is now the other way around, when an agent with a certain opinion (e.g., \emph{green})  meets two neighbours of a different opinion (e.g., \emph{non-green}) the change probability is higher the more agents in his block do not show this behaviour (i.e., the more show a \emph{green} behaviour). 

The exploration probability  $e$ should be small compared to $\tau$. 
Since an agent first has to interact with two agents of a different state at the same time in order to have a higher chance for switching its state, it is hard for the dynamics to escape from a situation where agents in a block have converged. As a consequence, the dynamics is metastable. The exploration probability only offers a small chance for an agent to change its state.

The dynamics of the whole population can again be viewed as a Markov chain $(\mathbf{X}_t)_{t\in\Z}$ on the state space  $\X = \{0,1\}^{2\times N}$, where we denote the population state at time~$t$ by $\mathbf{X}_t = (\mathbf{B}_t, \mathbf{O}_t) =(B^i_t, O^i_t)_{i=1}^N $. Requiring  $0<e, \tau<1$  ensures  that the Markov chain is irreducible and aperiodic.

\paragraph{Resulting dynamics.}
\begin{figure}
     \centering
     \begin{subfigure}[b]{0.58\textwidth}
         \centering
         \includegraphics[width=\textwidth]{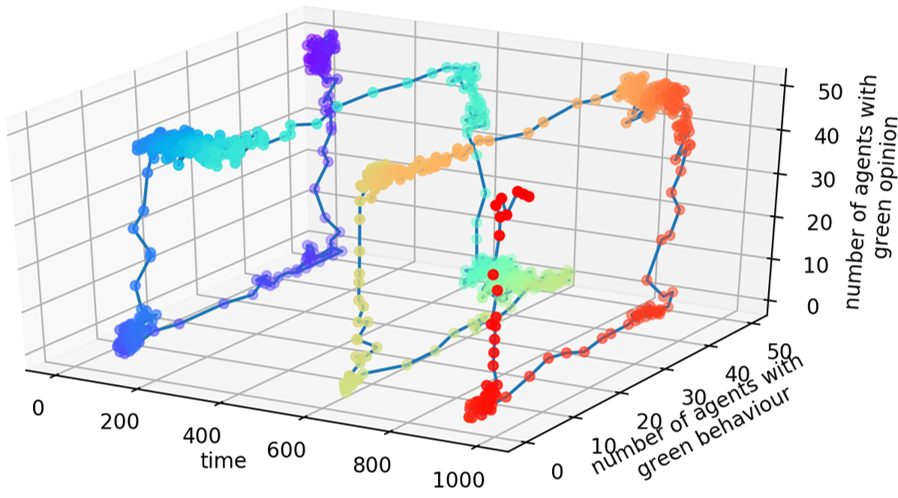}
         \caption{}
     \end{subfigure}
     \begin{subfigure}[b]{0.41\textwidth}
         \centering
         \includegraphics[width=\textwidth]{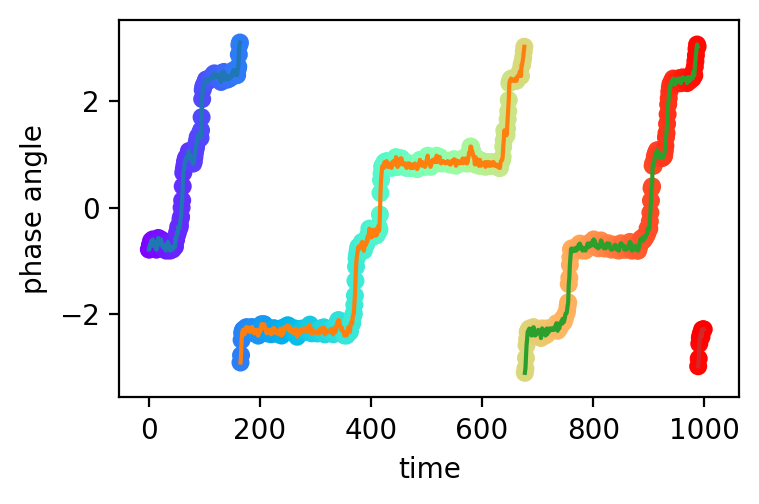}
         \caption{}
     \end{subfigure}

        \caption{Realization of the complex contagion dynamics for one complete network of $50$ interacting agents. (a) The dynamics is strongly cyclic in the plane spanned by the two coordinates ``number of agents with green opinion" and ``number of agents with green behaviour", (b) therefore it can   also be visualized by  plotting the clockwise-angle in the coordinate plane, i.e., the phase. The model parameters are $b,c,f,g=0.3$, $e=0.02$, $\tau=0.99$.}
        \label{fig:complex_complete}

\end{figure}
If we consider a fully-connected  population, in other words a complete network, then the dynamics cycles in one direction through the four possible metastable regions where the large majority of agents share the same opinion and display the same behaviour (either the one aligned with the shared opinion or the opposite one), see Figure~\ref{fig:complex_complete}~(a). Since the angle of rotation in the plane contains all information about the dynamics, we can essentially reduce the plot to~\ref{fig:complex_complete}~(b), where we show how the phase varies in time.

For this model  we are in the end interested in the possible transitions from a majority of agents with \emph{non-green} opinion and behaviour to a majority of agents with \emph{green} opinion and behaviour, which is a succession of several transitions between metastable states.

\begin{figure}
     \centering

     \parbox{0.5\textwidth}{
     \centering
     \begin{subfigure}[b]{0.5\textwidth}
         \centering
         \includegraphics[width=\textwidth]{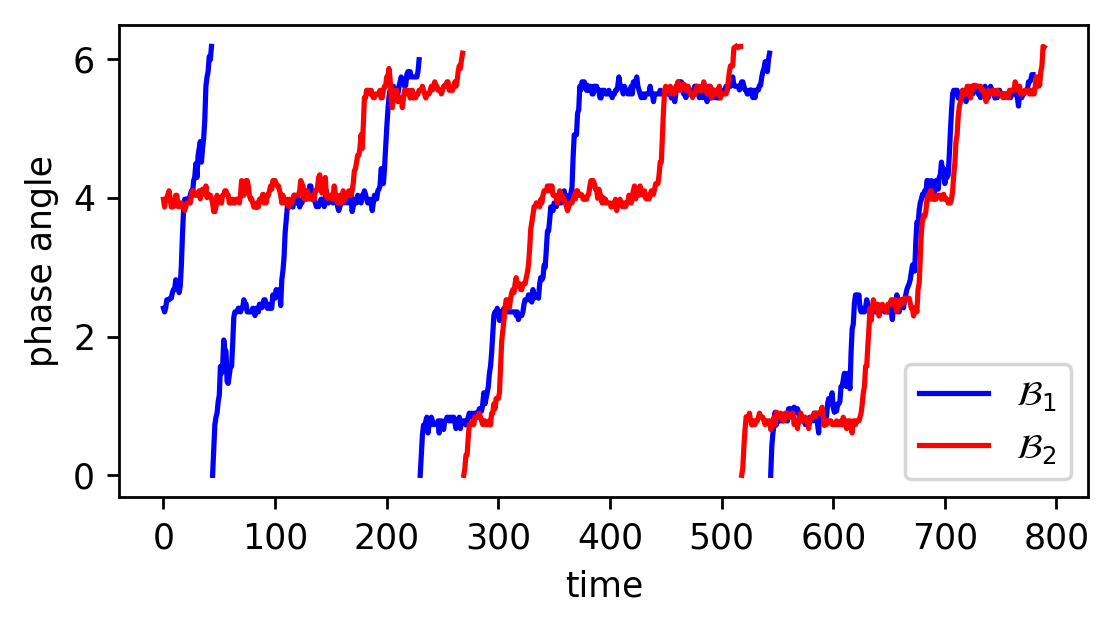}
         \caption{}
     \end{subfigure}
     
     \begin{subfigure}[c]{0.35\textwidth}
         \centering
         \includegraphics[width=\textwidth]{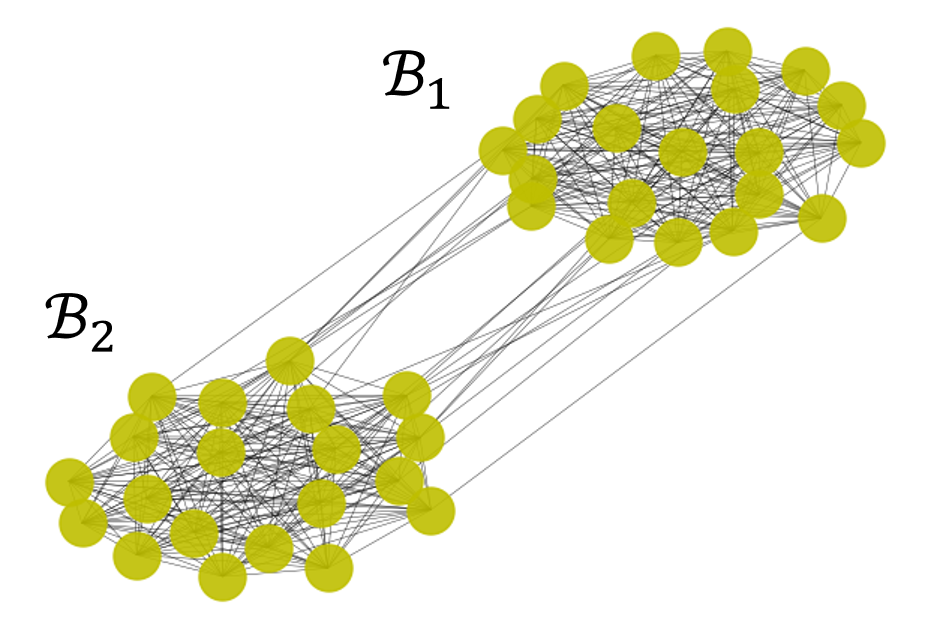}
         \caption{}
     \end{subfigure}}
          \begin{subfigure}[c]{0.47\textwidth}
         \centering
         \includegraphics[width=\textwidth]{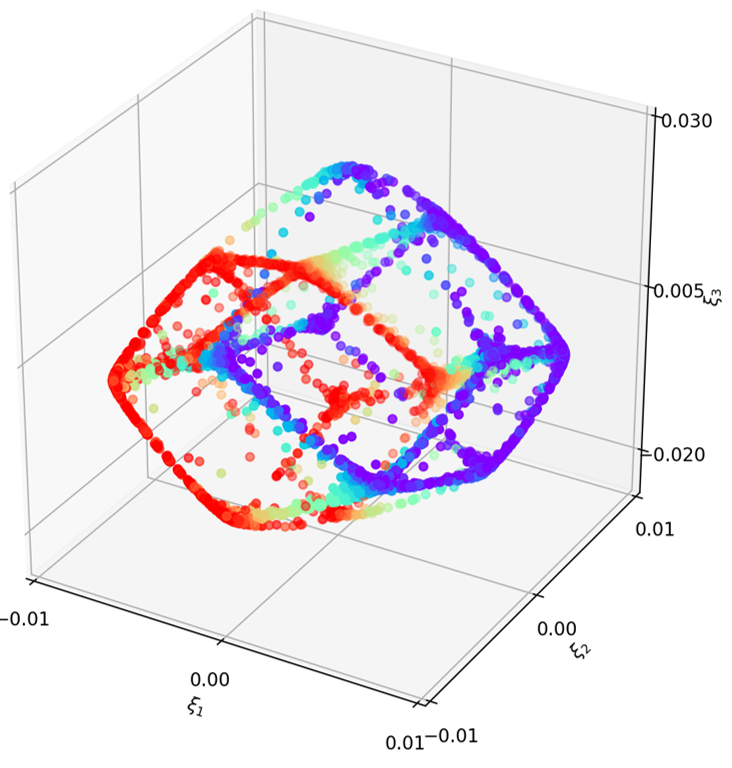}
         \caption{}
     \end{subfigure}

     \begin{subfigure}[b]{1\textwidth}
         \centering
         \includegraphics[width=\textwidth]{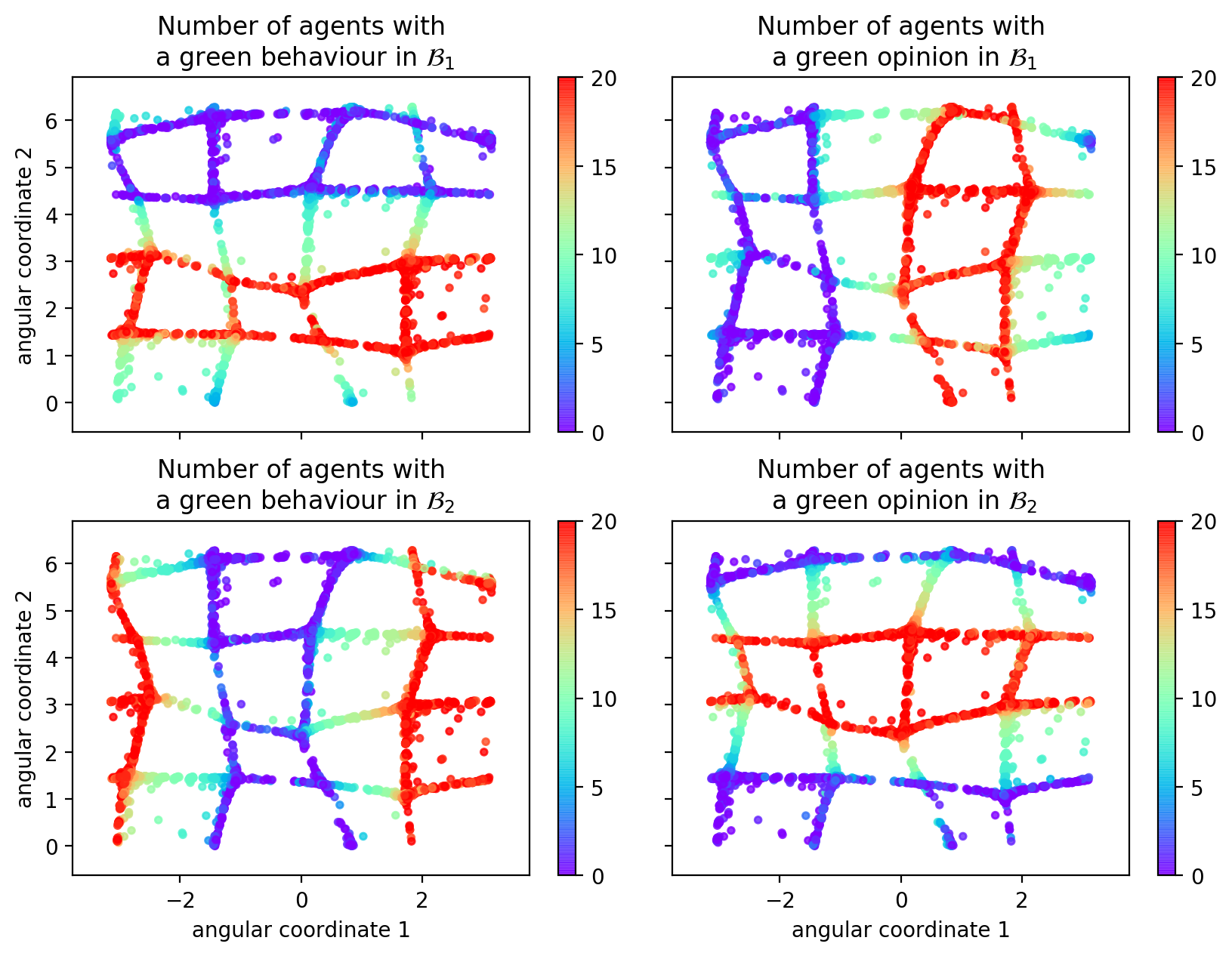}
         \caption{}
     \end{subfigure}

        \caption{Dynamics of the complex contagion model with two blocks as in \emph{Example~3}: (a) Realization showing the oscillatory dynamics of each block by plotting their phase, (b) network, (c) projection into the first three Diffusion Maps coordinates of the data set, the scale parameter $\epsilon = 0.1$.
        (d) Two angular coordinates extracted from the Diffusion Maps projection, colored according to the states of agents in the two blocks.}
        \label{fig:complex_2cluster}
\end{figure}

\textit{Example 3:} As an example throughout the paper, we consider a slightly larger population of $40$ agents split into two blocks, see Figure~\ref{fig:complex_2cluster}~(a)  for an example realization and Figure~\ref{fig:complex_2cluster}~(b) for the network.
As model parameters we set $b = c  = f = g = 0.7$, $e = 0.02$ and $\tau=0.99$. The internal wiring probability in each block is $W_{11}=W_{22}=1$, thus both blocks are complete.
An edge wiring probability of $W_{12}=W_{21}= 0.055$  between the two blocks ensures that the two blocks are mostly synchronized but still behave separately. The two blocks can also be viewed as two coupled oscillators where the coupling strength between oscillators is given by the number of connections between the blocks.

\begin{SCfigure}
    \centering
    \includegraphics[width=0.5\textwidth]{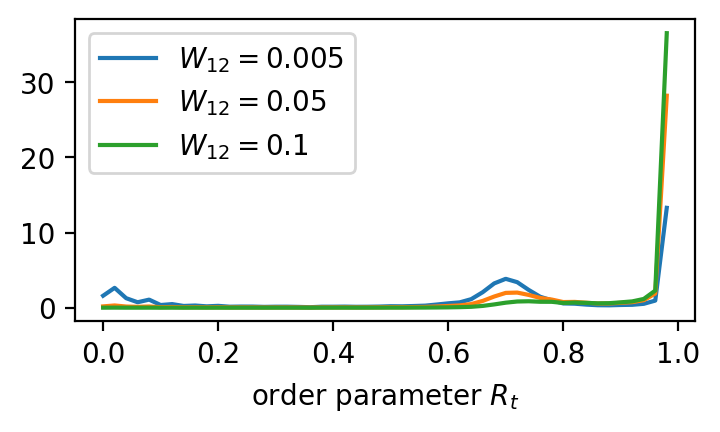}
    \caption{Distribution over time of the order parameter $R_t$,    Equation~\eqref{eq:orderparameter}, for the complex contagion model with different edge wiring probabilities $W_{12}$ between the two blocks.}
    \label{fig:orderparameter}
\end{SCfigure}

In Figure~\ref{fig:orderparameter} we study the distribution of the order parameter 
\begin{equation} 
R_t = \Big|\sum_{j=1}^2 \exp( \mathrm{i}\,  \theta^j_t) \Big|,
\label{eq:orderparameter}
\end{equation}
over time, which is a basic measure of synchronization between coupled oscillators~\cite{pikovsky2003synchronization, arenas2008synchronization}. Here, $\mathrm{i}$ denotes the imaginary unit. In our case we compare the level of synchronization for different edge probabilities $W_{12}$ between the two blocks and $\theta^j_t$ is the angle in the plane of ``number of agents with green opinion" vs ``number of agents with green behaviour" in block $j$ at time~$t$ as measured from the center point~$(20,20)$.
By placing every oscillator according to its phase on the unit circle, the order parameter measures the distance of the average of the positions on the unit circle from the origin. Thus when all oscillators are evenly spread out on the unit circle, $R_t$ is close to $0$, while when all oscillators are on the same spot, $R_t$ is $1$.
The results in Figure~\ref{fig:orderparameter} confirm that for our chosen edge wiring probability the two blocks are  synchronized most of the time. 

\section{Collective variables and reduced dynamics}\label{sec:coll_var}
One difficulty when analysing agent-based models lies in their high dimensionality. 
The size of the state space grows exponentially with the number of agents, 
and usually one is interested in studying a rather large population of agents.
If the system state resides most of the time in the vicinity of some low-dimensional manifold, 
then we can search for \emph{collective variables}, also called \emph{reaction coordinates} or 
\emph{order parameters}, 
$$\xi: \X \rightarrow \R^d$$ 
that allow an approximate description of the actual system's dynamics in a ``reduced''  state space with much lower dimension~$d$ than that of the original agent state space~$\X$. The reduced model approximately reproduces the emergent collective behaviour of the full ABM.
The reduction  allows us to better understand the structure of the dynamics and  eases numerical computations.

Fortunately, the dynamics of many ABMs have a low intrinsic dimension. On the one hand, groups of people in many social situations are behaving rather collectively and are influenced by their peers, e.g., through copying and imitating their peers or the opposite, being repelled from their neighbours' behaviour. On the other hand, real-world social networks are often highly modular, i.e., contain many communities, as well as being scale-free, i.e., having a few nodes with very high degree, which additionally encourages  coherent behaviour within sub-populations.  

When agents are rather homogeneous or identical, one can usually guess suitable collective variables based on an intuition about the system's dominant feedbacks. Moreover, if the collective variables are a ``simple" function of the agent state space, e.g., the number of agents that are in each of the different possible states, one can analytically derive mean-field approximations that take the form of coupled  ODEs or SDEs~\cite{porter2016dynamical,NiemannWinkelmannWolfetal.2020} and whose continuous-time formulation simplifies an analytical treatment even further. In our two example ABMs, we have identical agents which are however heterogeneous due to their different positions in the  network. This makes it more complicated to guess collective variables  and also deteriorates the approximation quality of rather straightforward mean-field approximations. In many ABMs there are further forms of heterogeneity, such as varying interaction parameters.

We therefore seek an automated way of finding collective variables $\xi$, which should allow us to represent the dominant model behaviour, as well as all the dominant and important transition pathways between the metastable regions in state space. We have selected here TPT as a very promising approach for studying transition dynamics. To be able to apply TPT to the resulting reduced model, we must also make sure that the projections of these metastable regions onto the low-dimensional manifold spanned by the sought collective variables are well separated from each other.

A similar task is performed by nonlinear manifold learning approaches. Given a cloud of sampled data points, $\mathbb{D}$, they try to parameterize the nonlinear manifold from which the data has been sampled.
In our case we want to parameterize the manifold close to which a large share of the observed system states, $\mathbb{D}=\{\mathbf{x}_1,\dots,\mathbf{x}_M \}$, lies once the system has become stationary.
We will use the dominant, real-valued coordinates produced by the Diffusion Maps algorithm~\cite{coifman2006diffusion,koltai2020diffusion} as collective variables (introduced in Section~\ref{sec:dmaps}).   
Diffusion Maps have already been applied for finding collective variables of ABMs~\cite{marschler2014coarse,liu2014coarse}, and for data-driven computation of dynamical quantities~\cite{thiede2019galerkin} that we will use to characterise tipping in Section~\ref{sec:tipping}.
They further benefit from being robust against noise, being computationally not too expensive, and the existence of an out-of-sample extension, with which one can interpolate the non-linear coordinates for data points not contained in~$\mathbb{D}$.

As the ABM dynamics is described by the Markov chain's large microscopic transition matrix, it will be suitable to also describe the identified collective variables' time evolution by a similar but much smaller macroscopic transition matrix. In order to find this reduced transition matrix  on the reduced state space $\xi[\X]$, we will  discretise the projected space 
and perform a Monte-Carlo estimation of the transition matrix, see Section~\ref{sec:proj_dyn}.

\subsection{Diffusion Maps}\label{sec:dmaps}
We want to apply Diffusion Maps~\cite{coifman2006diffusion,koltai2020diffusion} to a set of data points  \mbox{$\mathbb{D} = \{\mathbf{x}_1,\dots,\mathbf{x}_M \}\subset \X$} 
with the goal of finding a low-dimensional projection \mbox{$\xi:\X \rightarrow  \R^d$} of the given sample. 
The dimension $d$ should be small compared to the dimension of the original space~$\X$. 
For this to work well, we assume that the data points approximately lie on a $d-$dimensional manifold $\mathcal{M}$ embedded in the high-dimensional space~$\X$.  

The general idea of Diffusion Maps is to define a random walk on the data points~$\mathbb{D}$, where the transition probability between similar or near points  is high and between far points is close to zero. The random walk traverses the manifold and only follows its intrinsic structure, since the distances between near points in the original space are a good approximation to the local distances on the manifold. The dominant eigenvectors of the resulting transition matrix scaled by the corresponding eigenvalues  can then be used as a nonlinear projection.

The  transition matrix on the data points $\mathbb{D}$ is constructed as follows:
\begin{enumerate}
\item Choose a kernel $\ke (\mathbf{x},\mathbf{y}) = h\left(\frac{d(\mathbf{x},\mathbf{y})^2}{\epsilon}\right)$ that describes the similarity of two data points, for example the popular Gaussian kernel given by $h(z) = \exp(-z)$. Moreover one has to set the scale parameter $\epsilon>0$, e.g., by using the heuristic from~\cite{berry2016variable, koltai2020diffusion}, and a distance function $d(\cdot, \cdot)$ that is suitable for the data set, e.g., the Euclidean or Mahalanobis metric. 
\item Letting $q^\epsilon (\mathbf{x}_i) = \sum_{m=1}^M \ke(\mathbf{x}_i,\mathbf{x}_m)$, we form the new anisotropic kernel 
$$\ka (\mathbf{x}_i, \mathbf{x}_j) = \frac{k^\epsilon (\mathbf{x}_i, \mathbf{x}_j)}{q^{\epsilon }(\mathbf{x}_i)\, q^\epsilon(\mathbf{x}_j)}, $$
which has some desirable properties compared to~$\ke$, for more details see~\cite{coifman2006diffusion}.
\item Applying row-normalization by 
 $d^{\epsilon}(\mathbf{x}_i) = \sum_{m=1}^M \ka(\mathbf{x}_i,\mathbf{x}_m)$,
we arrive at  the  transition  matrix
 $$P^{\epsilon}(\mathbf{x}_i,\mathbf{x}_j)= \frac{\ka (\mathbf{x}_i,\mathbf{x}_j) }{d^{\epsilon}(\mathbf{x}_i)}.$$  
\end{enumerate}
The  matrix $P^{\epsilon}$ can be interpreted as the normalized Laplacian of a weighted undirected graph whose weights correspond to the anisotropic kernel~$\ka$. As such it is reversible with respect to the stationary distribution $\pi(\mathbf{x}_i) =\frac{d^{\epsilon}(\mathbf{x}_i)}{\sum_j d^{\epsilon}(\mathbf{x}_j)}.$

The right eigenpairs $(\lambda_j, \psi_j),\, j=0,\dots,M-1$ of  $P^{\epsilon}$ contain information about the  geometric structure of $\mathbb{D}$ at different scales and are real-valued due to $P^{\epsilon}$ being reversible.
We order the eigenpairs by decreasing magnitude of their eigenvalues. Then the leading eigenvectors, i.e., with the largest eigenvalues in magnitude, scaled by their corresponding eigenvalue, are a good projection of the large-scale structures in the data
 $$\xi(\mathbf{x}_i) = (\xi_{1,i}, ..., \xi_{d,i}) =  (\lambda_1 \, (\psi_1)_i, ..., \lambda_d \, (\psi_d)_i) \in \R^d,$$ 
 where $(\psi_j)_i$ is the $i$th component of the $j$th eigenvector.
Since the eigenvector corresponding to the largest eigenvalue is just the 1-vector and contains no information, we exclude it from the projection.
As $d$ we usually choose the number of remaining eigenvalues above the spectral gap. 
The Euclidean distances in these coordinates approximately correspond to the local diffusion distances on the manifold.  
 
The computational cost of computing pair-wise distances and the eigenvectors of~$P^{\epsilon}$ becomes very expensive if not impossible for very large data sets. To circumvent that, one can sub-sample the data set, compute the diffusion matrix and eigenpairs only for the sub-sample and interpolate the computed eigenvectors at the remaining data points  with the help  of the out-of-sample extension~\cite{coifman2006geometric}. We refer the reader to~\cite{koltai2020diffusion} for an explanation of the extension. 
 
\subsection{Collective variables of the two ABMs}\label{sec:proj_results}
Next we show the results of using the dominant Diffusion Map coordinates as collective variables for our two ABMs. We used the Diffusion Maps implementation from~\cite{cmdtools} and applied the algorithm to a sample of $20,000$ population states  \mbox{$\mathbb{D}=\{\mathbf{x}_1,\dots,\mathbf{x}_{20,000} \}$}. As a kernel we used the Gaussian kernel  and computed the distance  $d(\mathbf{x}_i,\mathbf{x}_j)$ between two data points via the Hamming distance, which measures the distance between two binary strings as the number of entries where they differ and is therefore suitable for binary population vectors. Further, we estimated an appropriate scale parameter~$\epsilon$ using the heuristic from~\cite{berry2016variable}.
 
For the following examples it is still possible to guess good collective variables. But when agents cannot be easily assigned to a block, then it will not be so clear how to choose collective variables. Most real-world social networks have rather fuzzy and overlapping communities, since people are part of several groups at the same time.
 
\paragraph{Threshold model.}
Since we set up the model such that agents in each block are nearly indistinguishable, the obvious choice  for collective variables for this system is just the number of active agents in each block (or equivalently, inactive agents). 
So let us see how the Diffusion Maps algorithm projects the data. 
 
\textit{Example 1 continued:}
The projection into the dominant two coordinates can be found in Figure~\ref{fig:granovetter_2cluster}. The sample of $20,000$ population states are embedded into a square. The coloring of the data points indicates that the two orthogonal directions encode the number of active agents in each block. Moreover note that the first Diffusion Map coordinate encodes the total number of active agents in the population and the second refines this by splitting them into two blocks. The Diffusion Map coordinates are refining the structure of the manifold with each additional coordinate and are ordered by the scales they encode. 
Looking more closely, we can see that the projected groups of points (corresponding to a certain number of active agents in each block) consist of some substructures on a smaller scale. These substructures encode whether agent $6$ is active or not, and how many of agent $0$ and $4$ are active (see Figure~\ref{fig:granovetter_2cluster}). Higher-order Diffusion Maps coordinates, in this case the coordinate $\xi_4$,  also decode the information about the activity of agents $0,4$ and $6$ (not shown in the figure). 
We will later investigate the importance of agents $0$, $4$ and $6$ with respect to the dynamics.

Judging from the Diffusion Maps eigenvalues, the intrinsic dimension of the dynamics seems to coincide with the number of blocks, i.e.,~$d=2$. 

\textit{Example 2 continued:} 
The projection onto the two most dominant coordinates out of the four dominant ones can be found in Figure~\ref{fig:granovetter_4cluster}. Though it is not so easy to see from just the dominant two coordinates, the data set of $20,000$ samples from a long realization  are embedded into a four-dimensional hypercube, a \emph{tesseract}, whose corners correspond to the states where the majority of agents in a certain set of blocks are active and the others not. The edges of the hypercube are much less visible but also present. They are not visited that frequently, since they correspond to the rare transitions between metastable regions. For our computations later we will use all four Diffusion Map coordinates.

\emph{Remark:} Note that the Diffusion Maps algorithm approximately preserves the local distances on the manifold on which the dynamics takes place. In some situations, e.g., for visualization purposes, it might be of interest to find even lower-dimensional embeddings that do not necessarily preserve the local distances but are still non-overlapping. For instance the net of the $3-$D cube can be embedded into $2-$D without overlapping edges by a planar graph projection.

\paragraph{Complex contagion model.}
For this model one would guess that the number of agents with green opinion in each block and the number of agents with green behaviour in each block will constitute good collective variables. Or, to reduce it further since the block behave like coupled oscillators, one could try using only these oscillators' phase angles in the plane spanned by the number of agents with green behaviour and opinion in each block.

\textit{Example 3 continued:} 
In Figure~\ref{fig:complex_2cluster} we show the Diffusion Maps projection into the dominant three coordinates, though four coordinates are needed to describe the dominant dynamics as indicated by the number of dominant eigenvalues of~$P^{\epsilon}$. Still, the three dominant coordinates in the figure already visually indicate that the data are projected onto a 4-D hypercube.

Diffusion Maps is not designed to embed a circle into a $1-$D torus, only into a $2-$D Euclidean space. Similarly, Diffusion Maps cannot embed the tesseract onto a $2-$D torus.
Here we will try to further post-process the embedding of $\mathbb{D}$ into $\R^4$ and project the net of the tesseract onto a 2-D torus. The tesseract can be projected onto the $2-$D torus without crossing edges. We choose two two-dimensional planes in $\R^4$  that are orthogonally meeting in the center of the projected tesseract and such that when measuring the angles in these planes, we can untangle the net of the tesseract without edges crossing each other. See Figure~\ref{fig:complex_2cluster} for the resulting net of the tesseract on the $2-$D torus. All the computations will still be done using the four Diffusion Maps coordinates, the projection onto the 2-D torus is only for visualization purposes.

Even though the projection indicates that the dynamics essentially takes place on a tesseract, we yet cannot infer how the dynamics moves along the edges of the tesseract. We know the model is strongly non-reversible, so on the edges there will be a dominant direction of the probability flux.
 
\subsection{Estimating the dynamics on the projected space}\label{sec:proj_dyn}
In order to study the reduced dynamics and apply Transition Path Theory for Markov chains, we need its transition matrix. 
We will explain how one can estimate a low-dimensional transition matrix that describes the dynamics on the projected space~$\xi[\X]$ from simulation data of the ABM.  

We assume that we have sampled i.i.d.\ pairs of consecutive states $\smash{ (\mathbf{x}^k,\mathbf{y}^{k})_{k=1}^K }$ of the Markov chain, in our case of the ABM. This means that $\mathbf{x}^k$ is sampled from the stationary distribution $\pi$ and $\mathbf{y}^{k}$ is sampled from the conditional transition probabilities~$ \trans(\mathbf{x}^k,\cdot)$.
Their projection into collective variables is given by~$\smash{ (\xi(\mathbf{x}^k),\xi(\mathbf{y}^{k}))_{k=1}^K }$. 
After partitioning the projected state space into  $M$ Voronoi cells $\{V_1,\dots, V_M\}$, e.g., by using the K-Means clustering algorithm, we can estimate a transition matrix \mbox{$\trans^{\xi} (m,n) = \Prob(\xi(\mathbf{X}_{t+1}) \in V_n \mid   \xi(\mathbf{X}_t) \in V_m)$} on the state space $\St=\{1,\dots,M\}$ identified with the Voronoi cells. 
We estimate $\trans^{\xi}$ using \emph{Ulam's method} by counting the proportion of transitions that went from $V_m$ to $V_n$ within one time step~\cite{Ulam60,schutte2013metastability, metzner2009estimating}: 
$$\hat{\trans}^{\xi} (m,n) = K^{-1} \sum_{k=1}^K \1_{V_m}(\mathbf{x}^k)\, \1_{V_n}(\mathbf{y}^k).$$ 
Instead of using many one-step trajectories, one can also use one long ergodic trajectory and count the one-step transitions therein. 
But for systems with many metastable regions, the ergodic trajectory needs to be very long to correctly sample the stationary density and to attain a good estimate of the transition probabilities.

\section{Studying tipping}\label{sec:tipping}
When we are interested in the transitions from one  subset  \mbox{$A \subset \St$} to another subset~\mbox{$B \subset \St$}, we can study the ensemble of trajectory pieces that start in $A$, end in~$B$, and in between only pass states in $C:=\St\setminus(A\cup B)$,  the so-called \emph{reactive trajectories}. TPT is a framework to get statistical information, e.g., the rate, density, flux and mean duration, about the reactive trajectories~\cite{weinan2006towards, Vanden-Eijnden2006,metzner2009transition}. If $A$ and $B$ are chosen as two metastable sets, then TPT studies the noise-induced tipping    between metastable regions. But $A$ and $B$ can be any  meaningful sets between which one wants to study the transition dynamics. TPT becomes particularly useful when tipping is very uncertain and there is a multitude of pathways linking $A$ to $B$.

The forward and backward committors contain all the essential information about the possible future and past transitioning behaviour and generate the various statistics of the ensemble of reactive trajectories.
The \emph{forward committor} gives the probability to first reach $B$ not $A$ when starting in $x\in \mathbb{S}$
$$q^+(x) :=  \Prob(\tau_B^+(t) < \tau_A^+(t) \mid  X_t=x),$$
where the random variable $\tau_S^+(t):= \inf\{s \geq t: X_s \in S \}$ is the  first hitting time of the set  $S\subset \mathbb{S}$ with the convention~$\inf \emptyset := \infty$. 
The \emph{backward committor} gives the probability to have last visited $A$  not $B$ when currently in $x$,
$$q^-(x) :=   \Prob( \tau_A^-(t) >\tau_B^-(t) \mid  X_t=x),$$
where we denoted the last exit time of the set $S\subset \mathbb{S}$ by \mbox{$\tau_S^-(t):= \sup\{s \leq t: X_s \in S \}$}, $ \sup \emptyset := - \infty$.
Due to the assumption of a stationary  process, the committors are time-independent.

Since we are in a stochastic setting, where tipping is usually not certain, we cannot define an interesting tipping point as a point of no return.
Instead, the tipping points or edge states can be identified with the points where tipping to $B$ is as likely as going back to $A$, i.e., those states with $q^+(x)$ close to $1/2$.
Recently it has been argued that the forward committor is the relevant object to quantify the risk of tipping to a certain state in the future~\cite{lucente2019machine,finkel2021learning}.

In this section we study ABMs described by a discrete Markov chain $(X_t)_{t\in \Z}$ on the reduced state space $\St=\{1,\dots,M\}$ that are stationary, irreducible and aperiodic. The ABM dynamics on $\St$ is described by the transition matrix $P^\xi$. In the following we will for simplicity write $P$.

We want to study tipping in ABMs that is only due to the noise facilitating rare transitions. But TPT has also been extended to non-stationary dynamics~\cite{helfmann2020extending}, i.e., Markov chains that are defined through an initial distribution $\mu$ and a possibly time-varying transition matrix $P(t)$. In this framework one can for example study noise-induced tipping in combination with parameter-induced tipping for a fixed parameter realization $\theta(t)$  on which the transition matrix $P(\theta(t)))$ depends. Moreover one can study tipping in models that are not stationary, e.g., in absorbing Markov chains that model for example epidemic processes with two absorbing states: where the virus dies out and where the virus has managed to infect every agent. 

In this section we will first introduce TPT before applying it to our two ABMs and studying their tipping behaviour in depth.

\subsection{Transition Path Theory}
We are interested in characterizing the transitions from one subset of the state space  $A\subset \St$ to another $B\subset \St$ (both non-empty and disjoint) in an irreducible and aperiodic Markov chain $(X_t)_{t\in \Z}$ on a finite state space $\St$. 
The time-homogeneous transition probabilities between states $x$ and $y$ are given by \mbox{$P(x,y) = \Prob(X_{t+1} = y\mid X_t =x)$}. Since we assume the process to be stationary, the distribution for all $t\in \Z$ is given by the \emph{stationary probability distribution} $\pi$, the unique probability-vector solution to $\pi^\top P = \pi^\top$.

TPT provides statistics about these transitions linking $A$ to $B$  in a stationary Markov chain by making use of the information contained in the \emph{forward committor}~$q^+(x)$, the probability to first hit $B$ rather than $A$ when starting in~$x\in \St$, and the \emph{backward committor}~$q^-(x)$, the probability to have last come from $A$ not $B$ when in~$x$.
The \emph{forward committor function} $q^+$
uniquely solves  
\begin{equation*}\label{eq:q_f}  
\left\{ \begin{array}{rcll}
q^+(x) &=& \sum\limits_{y\in \St} \, \trans(x,y) \, q^+(y) & x \in \C   \\
    q^+(x)&=& 0& x \in A  \\
    q^+(x) &=& 1& x \in B,  \\
\end{array}\right.
\end{equation*}
while the \emph{backward committor function} $q^-$
 is the unique solution to the linear system
\begin{equation*}\label{eq:q_b} 
\left\{ \begin{array}{rcll}
q^-(x) &=& \sum\limits_{y\in \St} \,
\transback(x,y) \, q^-(y) &x \in \C   \\
    q^-(x) &=& 0&x \in B  \\
    q^-(x) &=& 1& x \in A  \\
\end{array}\right.
\end{equation*}  
where $\transback(x,y) = \Prob(X_{t-1} = y\mid X_t =x) = \frac{\pi(y)}{\pi(x)} P(y,x)$ are the backward-in-time transition probabilities.

Next, we will define some statistics of the ensemble of 
reactive trajectories. 
As a \emph{reactive trajectory} we consider trajectory snippets \mbox{$(x_t, x_{t+1}, \dots, x_{t+T})$} that start in~\mbox{$x_t \in A$}, end in \mbox{$x_{t+T} \in B$} and in-between only pass through  \mbox{$x_{t+1},\dots,  x_{t+T-1}\in \C$}. 
The excursion \mbox{$(x_{t+1}, \dots, x_{t+T-1})$} through the transition region $\C$ is called an \emph{inner reactive trajectory}. 
When on a reactive trajectory  it holds that both $\tau_A^-(t) >\tau_B^-(t)$ and $\tau_B^+(t+1) < \tau_A^+(t+1)$, while on an inner reactive trajectory we have $\tau_A^-(t) >\tau_B^-(t)$ and $ \tau_B^+(t) < \tau_A^+(t)$.
The  ensemble of reactive trajectories is the collection of all reactive trajectory pieces that can be pruned out from the ensemble of stationary trajectories. 

The \emph{distribution} of inner reactive trajectories is defined as 
$$\mu^{AB}(x) =  \Prob(X_t=x,\tau_A^-(t) >\tau_B^-(t), \tau_B^+(t) < \tau_A^+(t)) = q^-(x)\, q^+(x)  \, \pi(x)$$ 
and indicates where inner reactive trajectories spend most of their time, and thus also the bottlenecks during transitions. The function $\mu^{AB}$ is not normalized,  but can be normalized by dividing by $Z^{AB} = \Prob(\tau_A^-(t) >\tau_B^-(t), \tau_B^+(t)< \tau_A^+(t))$, the probability to be on an inner reactive trajectory  at time $t$. 

The \emph{current} or \emph{flux} of reactive trajectories $f^{AB}$ denotes the probability of a reactive trajectory to visit $x$ and $y$ consecutively: \begin{equation*}
\begin{split}
f^{AB}(x,y)&= \Prob(X_t=x, X_{t+1}=y, \tau_A^-(t) >\tau_B^-(t), \tau_B^+(t+1) < \tau_A^+(t+1)) \\
&= q^-(x) \,\pi(x) \, P(x,y)   \,q^+(y) .
\end{split}
\end{equation*}
Again, this function is not normalized w.r.t.\ the pair $(x,y)$ and gives only the share of probability current accounting for reactive trajectories from $A$ to $B$. While the usual  probability current $\Prob(X_t=x, X_{t+1}=y)$ sums to $1$, the reactive current sums to $\Prob(\tau_A^-(t) >\tau_B^-(t), \tau_B^+(t+1) < \tau_A^+(t+1))=:H^{AB} $, which is the probability to be on a reactive trajectory.

The reactive current into a state $x\in \C$ equals the reactive current out of that state. $A$ and $B$ act as a source and sink of the reactive current, and the total reactive current out of $A$ equals the reactive current into $B$: 
$$\sum_{x\in A, y\in\St} f^{AB}(x,y) = \sum_{x\in\St, y\in B} f^{AB}(x,y).$$ 
We denote this quantity by $k^{AB}$, it specifies the \emph{rate} of reactive trajectories, since it estimates the number of reactive trajectories that are started in $A$ per time step, or equivalently, that end in $B$ per time step.
Further, by dividing the probability to be reactive by the transition rate, we obtain the mean duration of an inner reactive trajectory, $$t^{AB} = \frac{Z^{AB}}{k^{AB}}.$$ 
For us, the reactive current is the most important quantity, since it reveals the multitude of transition pathways from $A$ to $B$ and their respective weight.
Often one is interested in cycle-erased reactive trajectories from $A$ to $B$, which are free from uninformative cycles and only contain the progressive parts from $A$ to $B$. 
In the case of reversible dynamics, the effective current, which  gives the net amount of reactive current from state $x$ to $y$:
$$f^+(x,y) := \max\{f^{AB}(x,y)-f^{AB}(y,x),0\}$$ 
reduces to 
$\pi(x) \trans(x,y)(q^+(y)-q^+(x))$ for $q^+(y)>q^+(x)$ and $0$ else, and is therefore free of cycles.\footnote{
If this were not the case, then the committor would have to strictly increase along a cycle for $f^+(x,y)$ to be positive, but this is impossible.}
For non-reversible processes, such as the ones we study, the effective current is however not guaranteed to be cycle-free and we have to take a different approach. Moreover, it might be interesting to study the cycles of reactive trajectories separately.

\begin{SCfigure}
    \centering
    \includegraphics[width=0.55\textwidth]{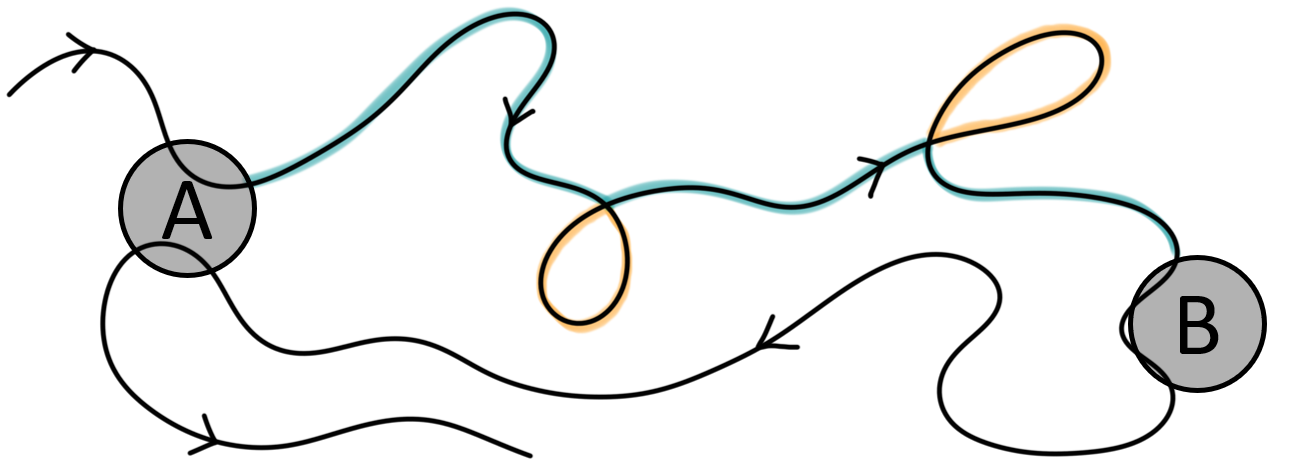}
    \caption{Splitting of a reactive trajectory into one productive path from $A$ to $B$ (in blue) and several unproductive cycles (yellow).}
    \label{fig:reac_paths}
\end{SCfigure}
\paragraph{Flux decomposition into productive and unproductive parts.}
In~\cite{banisch2015reactive} it is proposed to  decompose the reactive flux $f^{AB} = f^{P} + f^{U}$ into the flux $f^{P}$ coming from productive, cycle-free paths linking $A$ to $B$ and the flux $f^{U}$ from unproductive cycles. In Figure~\ref{fig:reac_paths} we sketch the productive piece of a reactive trajectory linking $A$ with~$B$ as well its two unproductive cycles.

Denote by $\Gamma^P$ the set of non-intersecting paths $\gamma = (x_1, \dots, x_s)$ that start in $x_1\in A$, end in $x_s \in B$, and go through the transition region $x_2,..,x_{s-1}\in \C$. By non-intersecting we mean that all of the traversed states $x_r$ are pairwise different, thus ensuring that the path is free of cycles. Further, all the visited edges $(x_r, x_{r+1})$ along the path need to have a positive transition probability~$P(x_r,x_{r+1})>0$.
By $\Gamma^U$ we denote the set of non-intersecting paths  $\gamma = (x_1, \dots, x_s)$ through the transition region~$x_r\in \C$ that are closed. Since the path is closed, it additionally contains the edge~$(x_s, x_1)$. Non-intersecting, closed paths are also called \emph{cycles}.
Note that self-cycles~$\gamma = (x)$, i.e. paths that go from~$x$ to~$x$, are also considered as cycles.

Now we are equipped to decompose the reactive current into the current from  cycle-free productive paths $\Gamma^P$ and the current from unproductive cycles $\Gamma^U$~\cite{banisch2015reactive}
    $$f^{AB}(x,y) = \underbrace{\sum_{\gamma \in \Gamma^P} w(\gamma) \, C^\gamma(x,y)}_{f^P} +  \underbrace{\sum_{\gamma \in \Gamma^U} w(\gamma) \, C^\gamma(x,y)}_{f^U}, $$
    where $C^\gamma$ is the incidence function of the path $\gamma$ 
    $$C^\gamma(x,y) =  
    \begin{cases}
    1, & \text{if } \gamma = (\dots,x,y,\dots) \\
    0, & \text{else}
   \end{cases} $$
   and $w(\gamma)$ encodes the path weight.  $w(\gamma)$ can be understood as an average along an infinitely long ergodic trajectory $(x_t)_{t\in\N}$ as follows
  \begin{equation}\label{eq:path_weight}
      w(\gamma) = \lim_{T \rightarrow \infty} \frac{N_T^\gamma}{T},
  \end{equation}  
  where $N_T^\gamma$ counts the number of times that $(x_t)_{t=1,\dots, T}$ passes through $\gamma$ while reactive. The edges of $\gamma$ have to be passed in the right order but excursions to one or more other cycles in between are allowed.\footnote{The original derivation in~\cite{banisch2015reactive} proceeds slightly differently. 
  They modify the reactive current to also include current from $B$ to $A$, and  then apply the \emph{stochastic cycle decomposition}~\cite{kalpazidou2007cycle,jiang2004mathematical}  for conserved currents to uniquely decompose the modified  current into cycles solely in $\C$ and cycles that contain an edge from $B$ to $A$. }

This decomposition into a productive and unproductive flux allows an interpretation of how much reactive trajectories are passing the different paths and cycles. 
The easiest way to numerically estimate this decomposition is as follows: 
   \begin{enumerate}
       \item Sample a long trajectory $(x_t)_{t=1,\dots, T}$ that contains sufficiently many transitions from $A$ to~$B$. Since we only need the reactive trajectory pieces for the computation of \eqref{eq:path_weight}  but correctly weighted by $H^{AB}$ compared to the non-reactive pieces, one can also use a transition matrix that only samples the reactive pieces of the trajectory correctly and maps all the non-reactive pieces to a single state~\cite{bowman2013introduction,cameron2014flows}. 
       \item Estimate $w(\gamma)$ by averaging along this sample trajectory~\cite{banisch2015cycle}. First prune out all the reactive pieces. Then for each reactive snippet iteratively cut out all the cycles  by going through the trajectory until for the first time a state is revisited, i.e., until we find $r$ such that $x_r = x_m$, $m<r$. Take out the cycle $(x_m, \dots x_{r-1}) = \gamma$ and increment  $N_T^\gamma$ by $1$. Repeat until from the reactive snippet only a cycle-free transition path $\gamma$ is left, increment $N_T^\gamma$ accordingly. Then move to the next reactive trajectory piece.
   \end{enumerate}
   
\paragraph{Flux aggregation.}
In order to assess the flux through certain subsets of the state space, e.g., through different channels or other regions of the state space, we can aggregate states together and compute the reaction current between aggregate region.
First we partition the state space into groups of states $L_s \subset \St$, in such a way that the disjoint union  of elements in $\{L_1,\dots,L_S\}$ is the whole state space $\St$ and that the boundaries of $A$ and $B$ are preserved. 
Then we can compute the \emph{reactive macro-current} $F^{AB}$ between the group of states $L_r$ and $L_s$  as follows~\cite{noe2009constructing}:
    $$F^{AB}(r,s) = \sum_{x \in L_r, y \in L_s} f^{AB}(x,y).$$
These macro-currents fulfill the same properties as the micro-currents, namely, the total flux out of $A$ equals the total flux into $B$, moreover the flux into a partition element~$L_r$ equals the flux out of that partition element~$L_r$.
We can also compute the effective macro-current~$F^+$ between partition elements from the reactive macro-current.

\subsection{Tipping analysis of the ABMs}
In this section we apply TPT to the two reduced ABMs in order to understand the possible tipping pathways.

When the two presented ABM dynamics are considered for modular agent networks, they have many metastabilities of different quality and strength. A multitude of transition paths exists that go from some region of interest $A$, via several stopovers in metastable regions, to some other region $B$. These transition pathways from $A$ to $B$ are forming a transition network.
We can understand the tipping dynamics also in terms of   tipping cascades among connected blocks: when one block tips,  it influences the probability of other connected blocks to also tip.  

\paragraph{Threshold model.}
For the threshold model we are interested in studying how the activity in collective behaviour spreads through the population. Therefore, we will set $A$ as all the states where a small proportion of agents is active and $B$ as all the states where the majority of agents in the population is active. 
Note that we have to be able to express $A$ and $B$ via the collective variables.
We used the TPT code from~\cite{pytpt}.

\begin{figure}
     \centering     
     \begin{subfigure}[b]{1\textwidth}
         \centering
         \includegraphics[width=\textwidth]{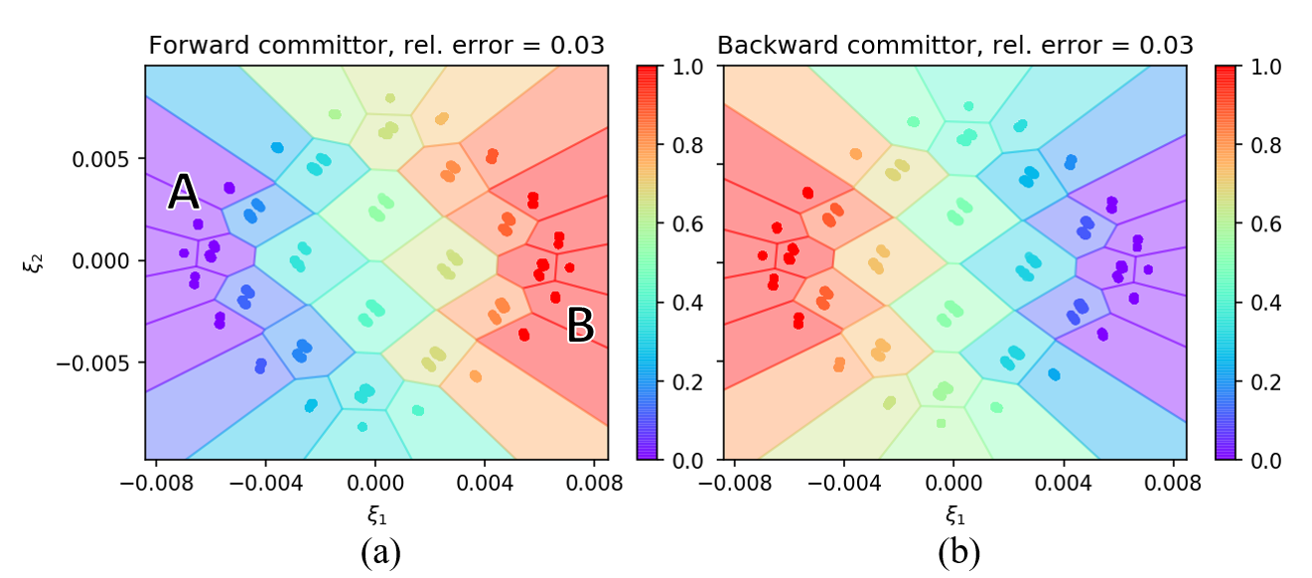}
     \end{subfigure}
     
     \begin{subfigure}[b]{1 \textwidth}
         \centering
         \includegraphics[width=\textwidth]{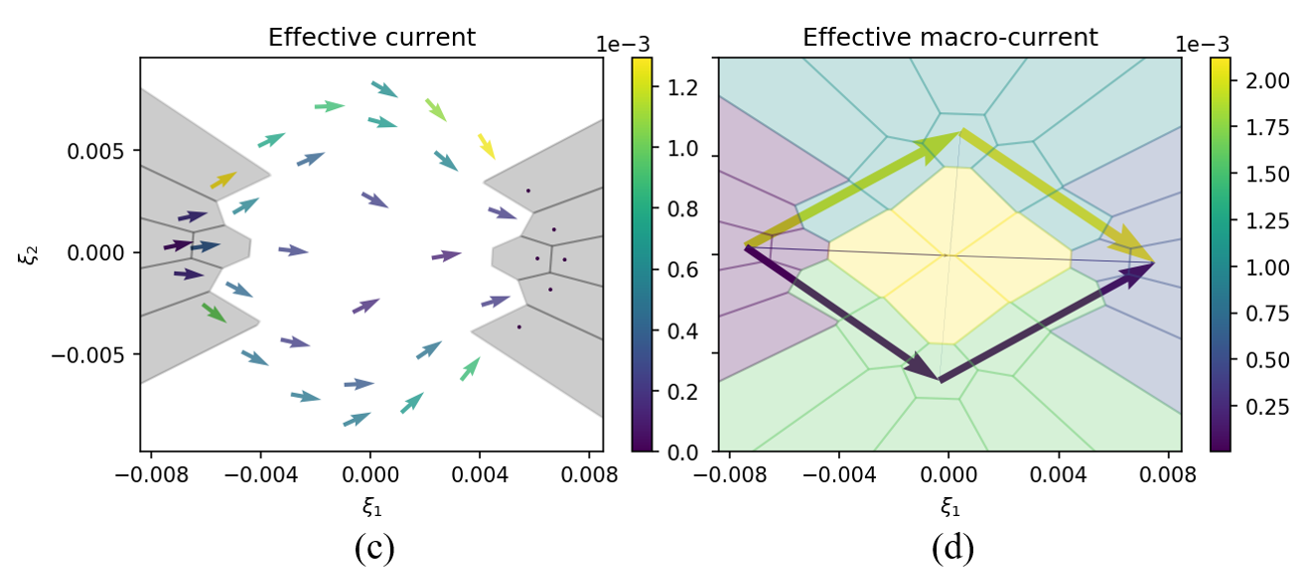}
     \end{subfigure}

     \begin{subfigure}[b]{1\textwidth}
         \centering
         \includegraphics[width=\textwidth]{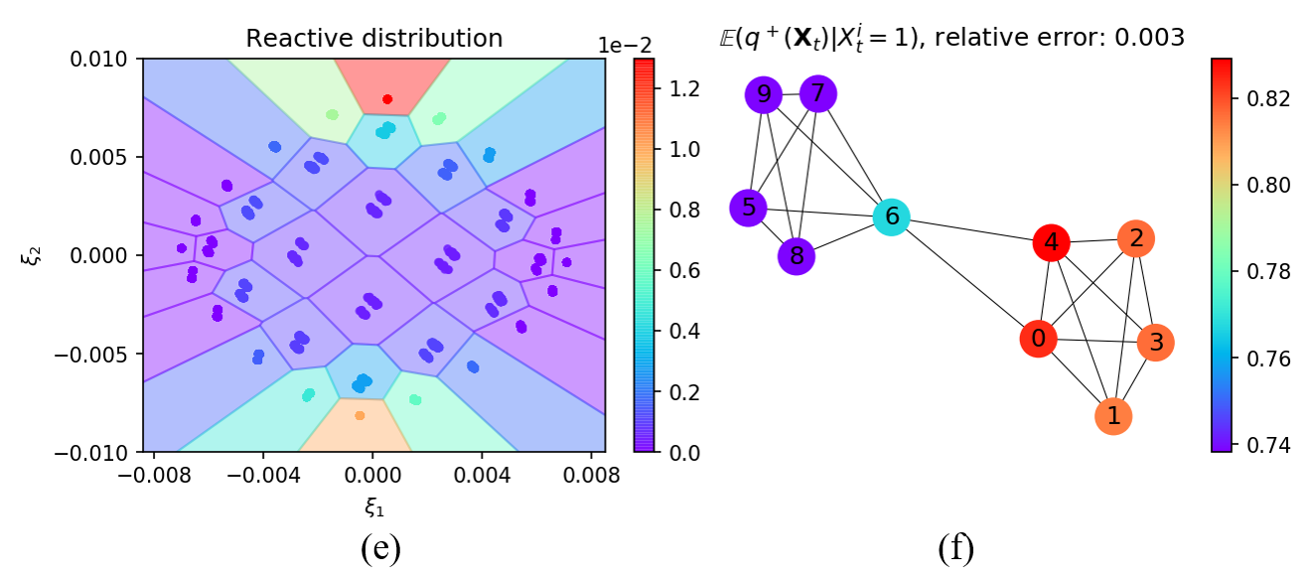}
     \end{subfigure}

        \caption{Tipping analysis for \emph{Example 1:}  (a), (b) Estimated committors on the discretized space, (c)  effective current, $A$ and $B$ are indicated by the two shaded areas, and (d)  effective macro-current through the three channels indicated in shaded blue, yellow and green. (e) Reactive distribution. (f) Agents as indicators of the overall tipping.}
        \label{fig:granovetter_2cluster_tpt}
\end{figure}
\emph{Example 1 continued}: 
We discretized the space spanned by $\xi$ into $36$ Voronoi cells, and estimated the transition matrix on this discrete space by using short trajectory snippets of total length $100,000$. 

The results of the tipping analysis between 
$A=\{\leq 2 \text{ agents are active}\}$ and $\qquad B=\{\geq 8 \text{ agents are active}\}$
are presented in Figure~\ref{fig:granovetter_2cluster_tpt}. In the panels (a) and (b) we show the committors on the reduced state space. We also computed the relative error between the estimated committors and the exact committors, 
which confirms that the collective variables represent the tipping dynamics well. 
The forward committor is not perfectly symmetric with respect to the two blocks: when block 1 has completely tipped but block 2 has not (these are the states around $\xi_1=0$ and $\xi_2>0.005$, compare with Figure~\ref{fig:granovetter_2cluster}~(c)), the forward committor is much higher than in the opposite scenario, when block 2 has tipped but block 1 not (the states around $\xi_1=0$ and $\xi_2<-0.005$). Also the reactive distribution is higher when block 1 has tipped and 2 has not. 
The transition rate amounts to $k^{AB} = 0.0039$ meaning that in a stationary  trajectory, a transition from $A$ to $B$ of duration $t^{AB}=18.85$ is started on average every $1/k^{AB} \approx 256$th time step.
From the effective current\footnote{The threshold model is only very slightly non-reversible, therefore we are not doing a flux-decomposition into cycles and productive parts and instead use the approximately cycle-free effective current.} in Figure~\ref{fig:granovetter_2cluster_tpt}~(e)  we can see that most of the transition flux from $A$ to $B$ goes along two pathways: 
\begin{itemize}
    \item[(I)] $A$ $\rightarrow$ agents in block 1 get active $\rightarrow$ agents in block 2 get active $\rightarrow$ $B$
    \item[(II)]  $A$ $\rightarrow$ agents in block 2 get active $\rightarrow$ agents in block $1$ get active $\rightarrow$ $B$.
\end{itemize}
In order to better compare the likelihood of both transition channels, we group the states of each channel together by hand, compare the coloring in Figure~\ref{fig:granovetter_2cluster_tpt}~(f), and compute the reactive macro-currents $F^{AB}$ and the effective macro-currents $F^{+}$ through these channels. 
Transitions along channel (I) contribute $53\%$ to $k^{AB}$, while channel (II) only contributes $41 \%$ to the rate.
We can now confirm that there is more effective current going through channel (I). The reason should lie in the asymmetry of the network between block $1$ and $2$: Agents $0$ and $4$ of block $1$ are both connected to agent $6$, see the network in Figure~\ref{fig:granovetter_2cluster}~(b). And from the ABM interaction rules we can deduce that the likelihood that agent $0$ and $4$ become active when agent $6$ is active is smaller than the likelihood that agent $6$ becomes active after agents $0$ and $4$.  These results also fit with the asymmetry in the committor: as soon as block 1 has become active, it is very likely that block 2 also becomes active.

To further study the role of each individual agent with respect to the overall tipping between $A$ and $B$, we consider the expected forward committor conditioned on agent $i$ being active.
When the forward committor is conditioned on agent $i$ being active,  the agents with the largest 
$$\E(q^+(X_t)\mid X^i_t=1) =: I^{AB}_i$$
are the best (individual-agent) \emph{indicators} that the overall tipping of the population will soon happen. When these agents are active, the system is the most likely to tip to $B$, thus one should especially consider these agents to access the tipping likelihood.

We can estimate  $I^{AB}_i$ by a Monte-Carlo approximation with a sufficiently long stationary ABM trajectory $(\mathbf{x}_t)_{t=1,\dots,T}$:
\begin{equation*}
\begin{split}
        I^{AB}_i 
        & = \frac{\E\left(q^+(X_t) \,  \1_{\{X^i_t=1\}}\right)}{\Prob \left(X^i_t=1\right)} 
        = \frac{\sum_{\mathbf{x}\in \X} q^+(\mathbf{x}) \,\1_{\{x^i=1\}}(\mathbf{x}) \, \pi(\mathbf{x}) }{\sum_{\mathbf{x} \in \X}  \1_{\{x^i=1\}}(\mathbf{x})\,\pi(\mathbf{x})} \\ 
        &\approx  \frac{\sum_{t=1}^T q^+(\xi(\mathbf{x}_t))\, \1_{\{x^i_t=1\}}(\mathbf{x}_t)}{\sum_{t=1}^T   \1_{\{x^i_t=1\}}(\mathbf{x}_t)}.
        \end{split}
\end{equation*}
From Figure~\ref{fig:granovetter_2cluster_tpt}  we can see that the agents from block 1 are the better tipping indicators. 
Moreover, agents $0$ and $4$ are the best indicators of tipping, while agent $6$ is the best indicator of block $2$. 
This is probably due to them being connected to the other block, thus increasing the tipping likelihood when they are active. 
One has to be careful in the interpretation of $I^{AB}$, it only shows us the correlations of the state of agent $i$ and the forward committor and not a causation, i.e.,  which agent has the largest individual impact on the overall tipping. \\

\begin{figure}
     \centering     
     \begin{subfigure}[b]{1\textwidth}
         \centering
         \includegraphics[width=\textwidth]{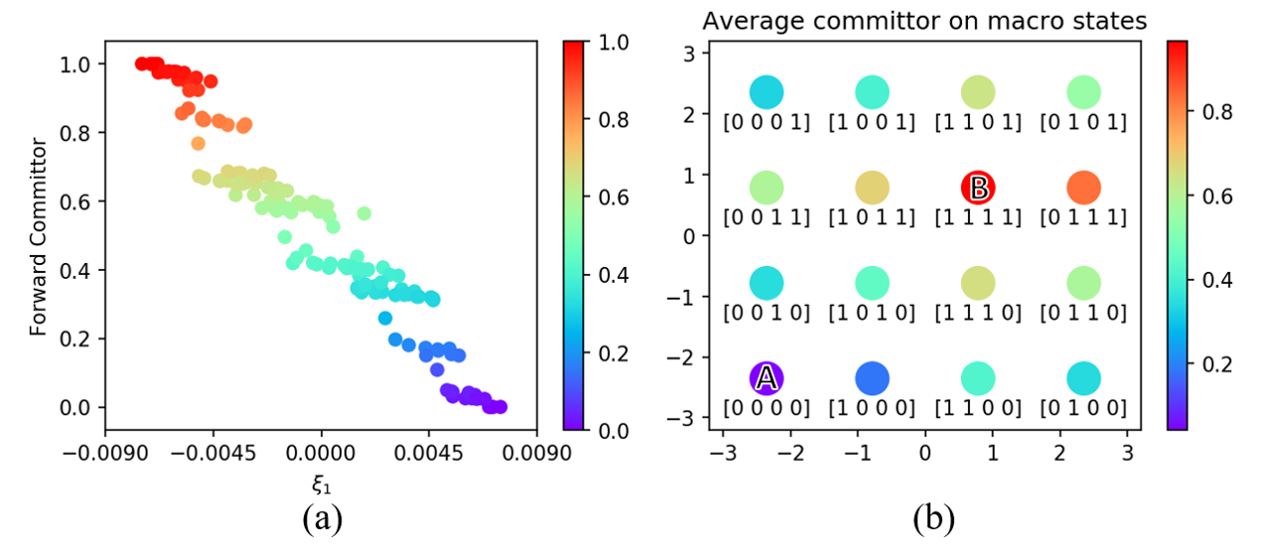}
     \end{subfigure}
 
     \begin{subfigure}[b]{1\textwidth}
         \centering
         \includegraphics[width=0.98\textwidth]{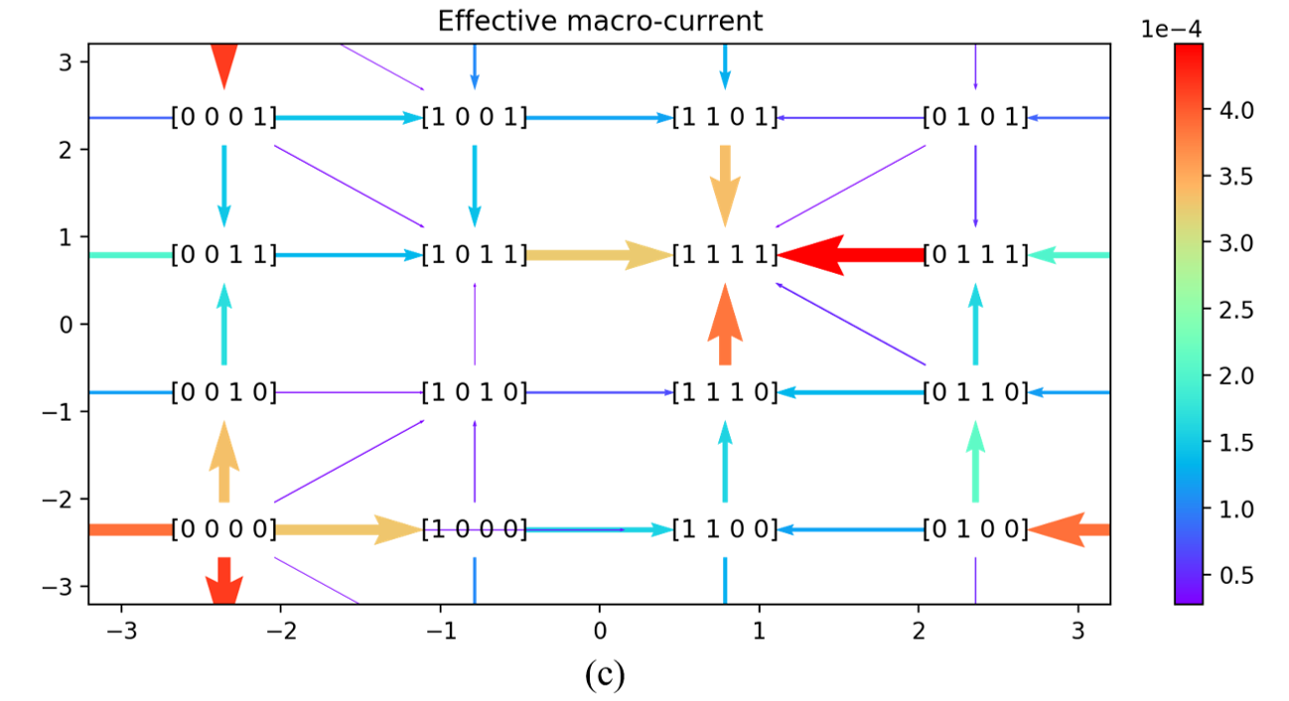}
     \end{subfigure}

     \begin{subfigure}[b]{1\textwidth}
         \centering
         \includegraphics[width=\textwidth]{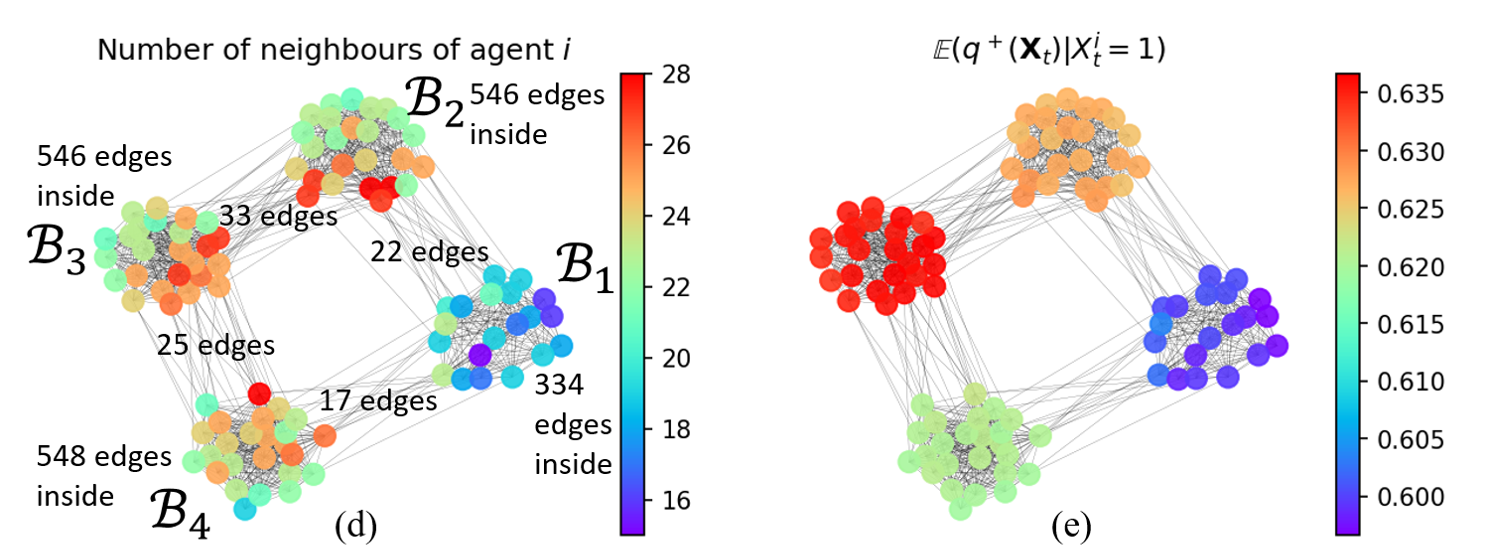}
     \end{subfigure}

        \caption{Tipping analysis of \emph{Example 2:}   (a) Forward committor against the dominant Diffusion Maps coordinate,  (b) mean forward committor on the macrostates that are placed on a torus. We denoted macrostates as a 4-D vector of $0$'s and $1$'s  decoding the majority activity in each of the four blocks, e.g.,  $[0,0,1,0]$   reads as majority of agents in block 1,2 and 4 are inactive and  majority in block 3 is active. (c)  Effective macro-current, the color and width of the arrow indicates the magnitude of the current. (d) Number of neighbours of each agent as well as the total number of connections inside and between blocks. (e) Agents as indicators of overall tipping. }
        \label{fig:granovetter_4cluster_tpt}
\end{figure}

\emph{Example 2 continued}:
After discretizing the projected state space into $150$ cells and estimating a transition matrix using $40,000,000$  trajectory snippets of total length, we show the tipping analysis for a population of four connected blocks in Figure~\ref{fig:granovetter_4cluster_tpt}. As $A$ we consider states where $\leq 25\%$ of agents are active, and as $B$ the states where $\geq 75\%$ are active.
The dominant Diffusion Maps coordinate encodes the number of agents that are active, and from  Figure~\ref{fig:granovetter_4cluster_tpt}~(a) we can see that along this coordinate the forward committor increases in distinct steps from $0$ to~$1$. 
Due to the faster decorrelation inside each metastable set, i.e., in the regions where agents in the same block are behaving conform, the  committor is  constant there.
From the committors we computed transition statistics, such as the average duration of reactive trajectories $t^{AB}=79.3$ and their frequency: In a long stationary trajectory, a transition from $A$ to $B$ is completed every $1/k^{AB}\approx 627$th time step. 

We again are interested in clustering states together in order to easier understand the transition dynamics and get a transition network. Since the system is much larger this time, we want to group cells together which are dynamically close by means of a clustering algorithm. Two popular approaches for a fuzzy dynamical clustering are PCCA+~\cite{roblitz2013fuzzy} and SEBA~\cite{froyland2019sparse}. We will use PCCA+ for non-reversible processes~\cite{conrad2016finding,fackeldey2018spectral,reuter2019generalized},  implemented in~\cite{cmdtools}, which takes the dominant real Schur vectors of the transition matrix and by a linear transformation maps them into a set of non-negative membership vectors that form a partition of unity and are as crisp as possible. Other optimization criteria also exist. By assigning each point to the block with the highest membership value, we can make the clustering crisp.
Since we expect the macrostates to be of the form where the majority of agents in each block is either inactive or active, we cluster the states into $2^4$ macrostates. These macrostates also correspond to the corners of the tesseract. The macrostates were then placed on a $2-$D torus such that the transition network can easily be visualized. 

In Figure~\ref{fig:granovetter_4cluster_tpt}~(b) and (c) we show the mean forward committor on the macrostates as well as the resulting transition network given by the effective macro-current. 
The macro-current is larger between macrostates where a neighbouring block tips than for a non-neighbouring block. Thus the dominant pathways from $A$ to $B$ are of the form of a tipping cascade from one block to its neighbours and then to their neighbours etc. 
The macro-current also indicates that it is most likely for block 4 to tip first and for block 1 to tip last. This can be explained as follows: Every agent in block $4$ has on average $21.92$ neighbours from the same block and $1.68$ neighbours from the other blocks. Compared to the other blocks, agents from block $4$ have the highest proportion of neighbours from the same block. Thus block 4 is the most independent block and therefore can change its activity most freely. The role of block 1 is also special. It is the smallest block with only $20$ agents and also the block where each agent has the largest proportion of extraneous neighbours. The role of block 1 is also reflected in the mean forward committor values: Out of all the macrostates, where only one block has tipped, the committor is the smallest when only block $1$ has tipped. This indicates that when block 1 has tipped, it easily tips back due to the strong influence from its neighbouring blocks. Moreover, out of all the macrostates where three blocks have tipped, the forward committor is the highest when block 1 is the still inactive block. 

For the network of four blocks we can study which agents are the best indicators of the overall tipping, see Figure~\ref{fig:granovetter_4cluster_tpt}~(e). We can immediately see that the values of $I^{AB}_i$ do not differ that much for the different agents, possibly due to the four blocks being of a rather similar size and similarly connected.  Still, block 3 seems to result in the highest expected forward committor when an agent of that block is active. Block 3 has the most connections to other blocks, and can therefore possibly exert the most influence on neighbouring blocks. This might explain why the expected forward committor is the largest when an agent from block 3 is active.
Moreover by comparing Figure~\ref{fig:granovetter_4cluster_tpt}~(d) with (e), there seems to be a very slight correlation of $I^{AB}_i$ with the number of neighbours an agent has.

\paragraph{Complex contagion model.}
In this model we are interested in analysing the tipping pathways between states where the majority has a non-green opinion and behaviour to states, where 
the majority has a green opinion and behaviour.

\begin{figure}
     \centering     

     \begin{subfigure}[b]{1\textwidth}
         \centering
         \includegraphics[width=\textwidth]{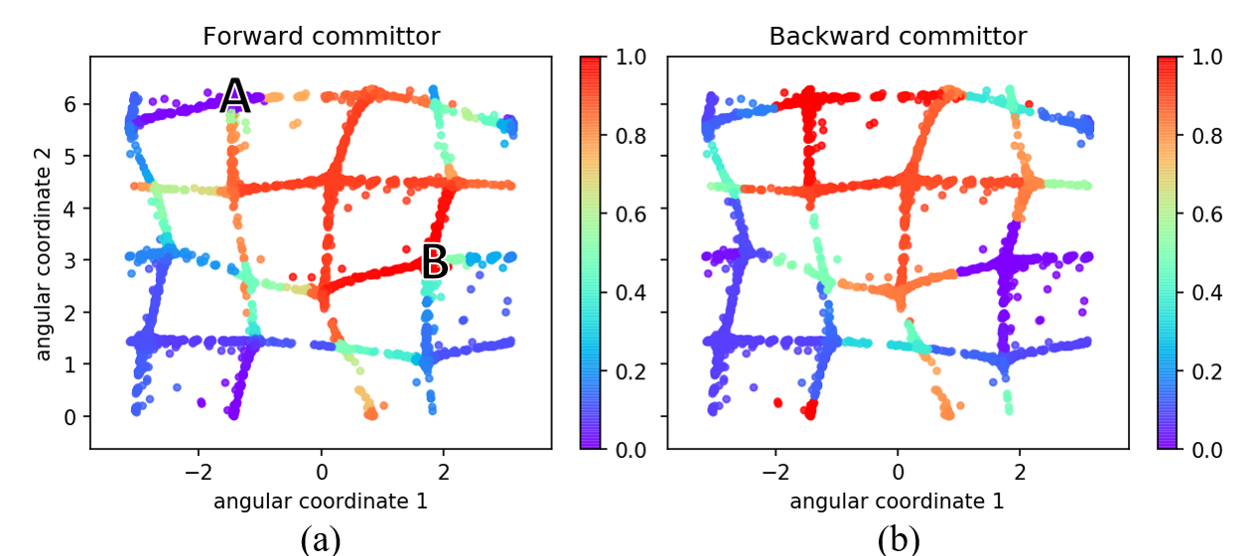}
     \end{subfigure}
     \begin{subfigure}[b]{1\textwidth}
         \centering
         \includegraphics[width=\textwidth]{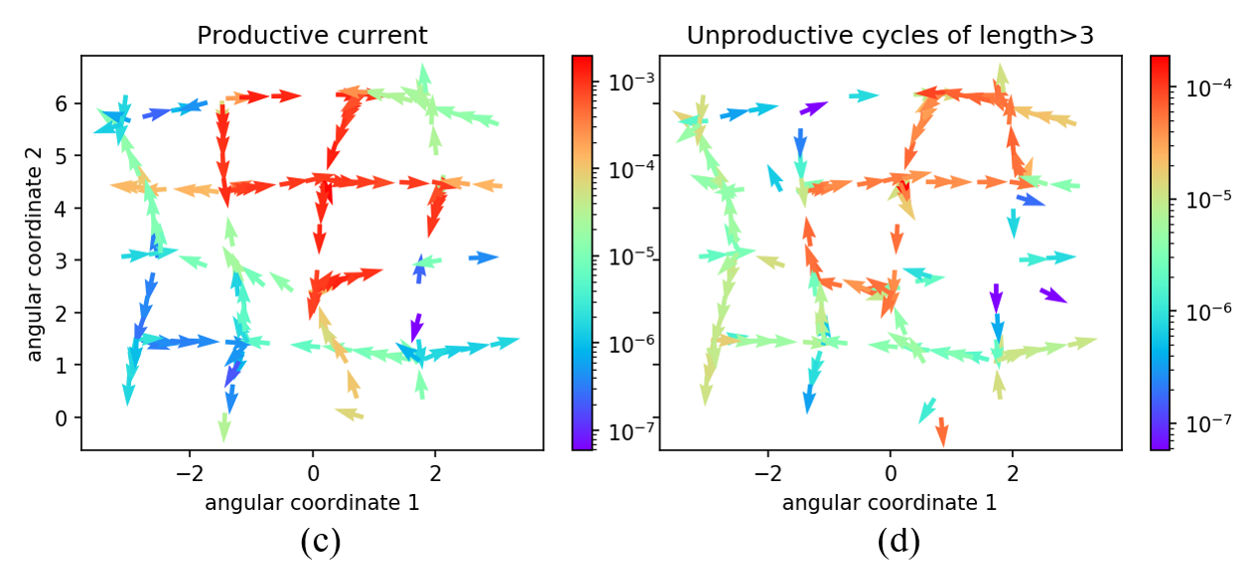}
     \end{subfigure}

        \caption{Tipping analysis of \emph{Example 3:} (a) Forward and (b) backward committor. (c)  Productive, cycle-free current from $A$ to $B$ (note the logarithmic colour scale). (d)  Unproductive current of  cycles  whose length is larger than~3. In order to get a clearer picture, we only plotted the flux produced by large cycles.}
        \label{fig:complex_2cluster_tpt}
\end{figure}

\emph{Example 3 continued}: 
We discretized the projected space into $150$ Voronoi cells and estimated the transition matrix on this space using $100,000$ short trajectory snippets.
The tipping analysis between the regions $$A=\{\leq 20\% \text{ of the population have a green opinion and behaviour}\},$$ $$B=\{\geq 80\% \text{ of the population have a green opinion and behaviour}\}$$  is shown in Figure~\ref{fig:complex_2cluster_tpt}.  
The two blocks behave as coupled oscillators that are mostly synchronized in a stationary regime. When the majority of agents in one block changes their behaviours or opinions, the other block will soon follow. Since the dynamics is moreover very cyclic, the forward committor is rather deterministic for many states, i.e.,  close to $0$ and $1$, see  Figure~\ref{fig:complex_2cluster_tpt}~(a). In order to better understand the projected states, compare with the coloring in Figure~\ref{fig:complex_2cluster}~(d). When the two blocks first change their opinions and then their behaviours from \emph{non-green} to \emph{green}, the forward committor is close to $1$ and when they change their opinions and behaviours back to \emph{non-green}, the forward committor is close to $0$. The backward committor is similarly very deterministic.
But there are also a few states where the committors are close to~$\frac{1}{2}$, so we will next look at the  current in order to understand the possible transition pathways from $A$ to~$B$.

Due to high non-reversibility, the effective flux is no longer cycle-free. Instead we can decompose the reactive flux into a productive, cycle-free  and an unproductive cyclic flux, see Figure~\ref{fig:complex_2cluster_tpt}~(c) and (d). 
From the decomposition we see that the dominant productive pathways are of the form:
\begin{itemize}
\item[(I)] $A$ $\rightarrow$ agents in one of the blocks change their opinion to \emph{green} $\rightarrow$ agents in the other block change their opinion to \emph{green} $\rightarrow$ agents in one of the blocks change their behaviour to \emph{green} $\rightarrow$ agents in the other block change their behaviour to \emph{green} $\rightarrow$ $B$,
\end{itemize}
while there are also some less likely productive paths:
\begin{itemize}
\item[(II)] $A$ $\rightarrow$ agents in one of the blocks change their opinion to \emph{green}  $\rightarrow$ agents in the same block change their behaviour to \emph{green}  $\rightarrow$ agents in the other block change their opinion to \emph{green} $\rightarrow$ agents in the other block change their behaviour to \emph{green}  $\rightarrow$ $B$.
\end{itemize}
The dominant unproductive cycles are of the general form:
\begin{itemize}
\item[(III)] 
Both blocks have a \emph{non-green} behaviour, majority of agents in block 1 (resp.~2) have a \emph{green} opinion $\rightarrow$ agents in  block 2 (resp.~1) change their opinion to \emph{green} $\rightarrow$  agents in block 2 (resp.~1) change their behaviour to \emph{green}  $\rightarrow$ agents in  block 2 (resp.~1) change their opinion back to \emph{non-green}  $\rightarrow$ then agents in block 2 (resp.~1) change their  behaviour back to \emph{non-green}.
\end{itemize}
In these unproductive cycles, one  block does a solo-cycle through the behaviour and opinion space. These are also common in coupled oscillators and called ``$2 \pi$ phase jumps"~\cite{pikovsky2003synchronization,arenas2008synchronization}.

By comparing the strength of the flux along the dominant productive paths (I) (around $9\times 10^{-4}-10^{-3}$) with the values of the current along the dominant unproductive cycles (III) (around $4\times 10^{-5}-6\times 10^{-5}$) in Figure~\ref{fig:complex_2cluster_tpt}~(c) and~(d), we can deduce that the pathways (I) are visited $15-25$ times as much as the dominant cyclic structure (III). 

Beyond the dominant pathways, we can give some general quantitative statements: Conditioned on being on a reactive trajectory, the probability to be on a productive path is $(H^{AB})^{-1} \sum_{x,y} f^P(x,y) = \frac{0.05}{0.285}=0.175$, while the probability to be on a cycle of length $>3$ is $(H^{AB})^{-1}  \sum_{x,y} f^{U,>3}(x,y)   = \frac{0.004}{0.285}=0.014$. The remaining conditional probability is attributed to cycles of length $\leq 3$.

\section{Conclusion}
In this paper we showed how to study noise-induced tipping pathways in high-dimensional, stationary models of heterogeneous agents. For complicated  agent-based models, 
analytically deriving reduced equations, e.g., ODEs or SDEs, is no longer possible or one has to accept large approximation errors. Here we instead used simulations of the  model in order to estimate a low-dimensional representation of the population states in terms of collective variables. 
In our two guiding models, agents are strongly affiliated with a subpopulation. Due to the local interaction rules,  those population states, where the individual agents in the same subpopulation agree in their actions or attitudes, are metastable. We found that the population states approximately lie on the skeleton of a hypercube that can be parametrized with just a few coordinates. The corners of the hypercube represent the metastable states while the edges make up the transition paths between metastable states. 
Thus the estimated reduced states can describe all the macro-scale patterns and large shifts and changes in the population of agents. 
In the two considered ABMs, tipping  between the two extreme metastable regimes in the system happened as a tipping cascade among connected subpopulations. 
We applied Transition Path Theory to quantify the tipping dynamics and could for instance uncover the dominant cascading pathways as well as possible loop-dynamics on the way from $A$ to $B$.

It is noteworthy to mention that TPT can quantify the tipping paths without relying on actual sampled transition paths. By estimating the transition matrix from short samples, the local information in the short samples can be combined to solve for the global committor functions. 

By studying which agent $i$ results in the highest expected forward committor conditioned on them being in  a certain state, $I^{AB}_i$,  we could assess which agents  in the network are the best indicators for tipping towards $B$.

Several open aspects and questions remain:
\begin{enumerate}
\item In order to better understand the quantity  $I^{AB}_i$, one could systematically study $I^{AB}_i$ for different  small networks, similar as in~\cite{holme2017three, holme2018epidemic}, as well as compare it to different centrality measures for network nodes.
\item By studying other general types of ABM dynamics or interactions on non-modular networks, could one find other generic forms of the low-dimensional manifold on which the population states concentrate?
\item Another prospect would be the study of more realistic ABMs or dynamics on real-world networks.
\end{enumerate}

\paragraph{Acknowledgements: }
We would like to thank Marc Wiedermann for discussions about   Granovetter's threshold model and Alexander Sikorski for helping with speeding-up the ABM simulation code and many discussions. 
We are grateful to Marvin Lücke for carefully reading the manuscript. 
Luzie Helfmann acknowledges support by the Deutsche Forschungsgemeinschaft (DFG, German Research
Foundation) under Germany´s Excellence Strategy – The Berlin Mathematics
Research Center MATH+ (EXC-2046/1, project ID: 390685689).
{
\small
\bibliographystyle{plain}
\bibliography{references}
}
\end{document}